\documentclass[prX,aps,amssymb,shownopacs,twocolumn,10pt,floatfix]{revtex4-1}
\usepackage{amsmath}
\usepackage{amssymb}
\usepackage{amsthm}
\usepackage{amsfonts}
\usepackage{listings}
\usepackage{easybmat}
\usepackage{enumerate}
\usepackage{latexsym}
\usepackage{hyperref}

\usepackage{psfrag}
\usepackage{color}
\usepackage{bm}
\usepackage[pdftex]{graphicx}
\usepackage{subfigure}

\def\ie{{\it i.e.},\ }
\def\eg{{\it e.g.}\ }

\def\ket#1{\left|#1 \right\rangle}
\def\bra#1{\left\langle #1 \right|}
\def\braket#1#2{\left\langle #1 | #2 \right\rangle}
\def\matrix22#1#2#3#4{\left(\begin{array}{cc}#1&#2\\#3&#4\end{array}\right)}

\usepackage{ulem}

\begin{document}

\title{Interacting bosons in topological optical flux lattices}

\author{A. Sterdyniak$^1$}
\author{B. Andrei Bernevig$^2$}
\author{Nigel R. Cooper$^{3}$}
\author{N. Regnault$^{2,4}$}
\affiliation{$^1$ Institute for Theoretical Physics, University of Innsbruck, A-6020 Innsbruck, Austria\\
$^2$ Department of Physics, Princeton University, Princeton, NJ 08544\\
$^3$ T.C.M. Group, Cavendish Laboratory, J.J. Thomson Avenue, Cambridge CB3 0HE, United Kingdom\\
$^4$ Laboratoire Pierre Aigrain, ENS-CNRS UMR 8551, Universit\'es P. et M. Curie and Paris-Diderot, 24, rue Lhomond, 75231 Paris Cedex 05,France}

\begin{abstract}

An interesting route to the realization of topological Chern bands in ultracold atomic gases is through the use of optical flux lattices. These models differ from the tight-binding real-space lattice models of Chern insulators that are conventionally studied in solid-state contexts. Instead, they involve the coherent coupling of internal atomic (spin) states, and can be viewed as tight-binding models in reciprocal space. By changing the form of the coupling and the number $N$ of internal spin states, they give rise to Chern bands with controllable Chern number and with nearly flat energy dispersion. We investigate in detail how interactions between bosons occupying these bands can lead to the emergence of fractional quantum Hall states, such as the Laughlin and Moore-Read states. In order to test the experimental realization of these phases, we study their stability with respect to band dispersion and band mixing. We also probe novel topological phases that emerge in these systems when the Chern number is greater than 1.
\end{abstract}

\date{\today}
\maketitle

\section{Introduction}\label{sec:intro}

Recent years have witnessed a surge of interest in variants of the fractional quantum Hall (FQH) problem, which depart from the original setting of interacting electrons in uniform magnetic field by replacing the electrons by interacting bosons and/or by introducing strong lattice effects\cite{Kol:1993p82,Srensen:2005p58,Palmer:2006p63,Hafezi:2007p67,Palmer-PhysRevA.78.013609,Moller:2009p184,Moller:2012p2646,Sterdyniak-PhysRevB.86.165314,Scaffidi-2014arXiv1407.1321S}. These generalizations are motivated both by new experimental settings where such questions emerge naturally (\eg in ultracold atomic gases, or materials with strong spin-orbit coupling), and also, in light of the many forms of topological insulator that are now understood to be possible, by the urgent search to understand the full range of possible {\it strongly correlated} topological phases. 

For the case of two-dimensional lattices, with ``Chern bands'' replacing the lowest Landau level, the generalized strongly correlated phases are referred to as fractional Chern insulators (FCIs)\cite{neupert-PhysRevLett.106.236804,sheng-natcommun.2.389,regnault-PhysRevX.1.021014,BERGHOLTZ-JModPhysB2013,Parameswaran2013816}. Most theoretical studies of FCIs have focused on tight-binding lattice models for the Chern bands, these being most readily related to electronic materials. Significant understanding has been reached in how to generate topological bands with controllable Chern number and with controllable dispersion, typically by introducing and tuning hopping parameters. However, in the context of ultracold gases -- where the full range of generalized fractional quantum Hall systems has great potential for experimental exploration -- such tight-binding models are not necessarily the most natural models to consider. Indeed, it has been shown that adaptations of schemes involving the Raman coupling of internal atomic states~\cite{dalibardreview} (as studied in experiments at NIST~\cite{spielmanfield}) to optical lattice geometries allows the formation of topological bands (including Chern bands) in lattices in which the atoms remain far from the tight-binding limit. Instead, these ``optical flux lattices'' (OFLs)~\cite{ofl,cooperdalibard} are best understood in reciprocal space~\cite{PhysRevLett.109.215302}. Design strategies allow significant control over the band topology, dispersion and Berry curvature distribution of these OFLs, by controlling the number of internal states and the laser couplings~\cite{PhysRevLett.109.215302}. As an example, a practical scheme has been proposed for Raman coupling of $N=3$ internal states of $^{87}$Rb~\cite{PhysRevLett.110.185301}, which leads to flat lowest band with Chern number $C=1$, and strongly correlated bosonic FQH phases including the Moore-Read\cite{Moore:1991p165} phase even for the (weak) two-body interactions expected in that experimental setting.

Different from a single Landau level, CIs can be used to generate an almost flat band with a Chern number $C>1$. Once strong interactions are turned on, it is expected\cite{Barkeshli-PhysRevX.2.031013} that these systems host new phases that are a generalization of the Halperin states~\cite{Halperin83} in the FQH with color-orbit couplings. Indeed, numerical evidence for such states has been obtained in various FCIs~\cite{Wang-PhysRevB.86.201101,Yang-PhysRevB.86.241112,Liu-PhysRevLett.109.186805,PhysRevB.87.205137,Wu-PhysRevLett.110.106802,Wu-2013arXiv1309.1698W}. Since OFLs allow to tune the Chern number of the lowest band while preserving its approximate flatness, these systems appear to be natural candidates to implement these phases.

In view of the very wide range of control of band topology and dispersion that is possible using OFLs, they provide a very interesting and adaptable framework in which to study FCIs. In this paper we explore the range of bosonic FCIs that emerge in these model systems, varying the number of internal states that are coupled $N$ and the Chern number of the lowest band $C$. The possibility to implement these models directly in experiment raises important practical issues. We test possible experimental realizations of these phases by studying their stability to the band dispersion relation and to band mixing. We also probe novel topological phases that emerge in these systems when the Chern number is greater than one.

The paper is arranged as follows. In Section~\ref{sec:OFL}, we describe the non-interacting model of the bosonic atoms with $N$ internal degrees of freedom trapped in the optical flux lattice, introduced in Ref.\onlinecite{PhysRevLett.109.215302}, that we consider for our numerical simulations. In Section~\ref{sec:FCI}, we give a brief overview of FCIs both for $C=1$ and $C>1$. We also give a brief introduction to the particle entanglement spectrum~\cite{sterdyniak-PhysRevLett.106.100405} that we use to probe the different states. We then discuss in Section~\ref{sec:numC1} the numerical results for interacting bosons in an optical flux lattice with Chern number $C=1$. In particular, we compute the value of the neutral gap above the Laughlin state both in the flat band approximation and in the presence of band mixing. We also present evidence for the emergence of a Moore-Read phase with two-body interactions and we discuss its stability. Finally in Section~\ref{sec:numclt1} we provide a detailed study of the emergence of Halperin-like states in optical flux lattice with Chern number $C=2$ and $C=3$, including their topological signature, neutral gap and stability.

\section{Optical Flux lattices}\label{sec:OFL}

In this paper, we consider bosonic atoms with $N$ internal degrees of freedom (e.g. spin states) subjected to the optical flux lattice described in Ref.~\onlinecite{PhysRevLett.109.215302}. The internal states are coupled by two-photon Raman transitions driven by laser beams that are arranged to form a periodic lattice. (We consider a uniform system, made finite by applying periodic boundary conditions commensurate with this lattice, as described below.) The one-body Hamiltonian reads
\begin{equation}	
\hat{H} = \frac{{\bm P}^2}{2M} \hat{\openone}_N + \hat{V}({\bm r})
\label{eq::OFL_Ham}
\end{equation}
where $\hat{\openone}_N$ is the $N \times N$ identity matrix. $\hat{V}({\bm r})$ describes the coupling between the internal atomic states induced by the laser beams and is given by \cite{PhysRevLett.109.215302}

\begin{widetext}
\begin{equation}
{\hat{V}({\bm r}) = 
-  V
\left(\begin{array}{ccccc}
2\cos({\bm r}\cdot {\bm \kappa}_3) & A_1+A_2e^{-iC\frac{\pi}{N}} & 0 & \ldots & A_1^*+A_2^*e^{iC\frac{\pi(2N-1)}{N}} \\
 A_1^*+A_2^*e^{iC\frac{\pi}{N}} & 2\cos({\bm r}\cdot {\bm \kappa}_3 -\frac{2C\pi}{N}) &  A_1+A_2e^{-iC\frac{3\pi}{N}} & \ldots & 0\\
 0 &  A_1^*+A_2^*e^{i\frac{C3\pi}{N}} & 2\cos({\bm r}\cdot {\bm \kappa}_3-C\frac{4\pi}{N})  & \ldots & 0\\
\vdots & \vdots & \vdots & \ddots & \vdots \\
 A_1+A_2e^{-i\frac{C\pi(2N-1)}{N}} & 0 & 0 & \ldots & 
 2\cos({\bm r}\cdot {\bm \kappa}_3 -C\frac{2\pi(N-1)}{N}) 
\end{array}\right)}
\label{eq:vn}
\end{equation}
\end{widetext}
where $A_j=\exp(-i{\bm r}\cdot{\bm \kappa}_j)$, ${\bm  \kappa}_1= (1,0)\kappa\, , {\bm \kappa}_2=\left(\frac{1}{2},\frac{\sqrt{3}}{2}\right)\kappa$ and ${\bm  \kappa}_3={\bm \kappa}_2-{\bm \kappa}_1$. $C$ is an integer and corresponds to the Chern number of the lowest band of this model. The energy scale (bandwidth) of this Hamiltonian is given by the recoil energy $E_R= \frac{\hbar^2\kappa^2}{2M}$. Here we do not discuss experimental implementations of this model, but note that Ref.~\onlinecite{PhysRevLett.110.185301} showed how to implement a closely related model for $N=3$, and that schemes for gauge fields using the coupling of $N>3$ levels have been suggested in the literature~\cite{PhysRevA.84.025602,PhysRevA.88.011601}.

This Hamiltonian is invariant to translations along ${\bm a'}_1=\frac{2\pi}{\kappa}(1,-1/\sqrt{3})$ and ${\bm a'}_2=\frac{4\pi}{\kappa\sqrt{3}}(0,1)$. Thus, it can be diagonalized in the plane waves basis $\ket{\alpha,{\bm k},{\bm G}}$ where $\alpha=0 \dots N-1$ denotes the particle internal state, ${\bm k}$ is the momentum in the first Brillouin zone and ${\bm G}$ is a vector of the reciprocal space spanned by ${\bm \kappa}_1$ and ${\bm \kappa}_3$. In real space the wave function is simply $\braket{{\bm r}}{\alpha,{\bm k},{\bm G}} = \ket{\alpha}e^{i {\bm r}\cdot ({\bm k + {\bm G})}}$.
In this basis, the non zero matrix elements of $\hat{V}$ are: 
\begin{eqnarray}
\bra{\alpha,{\bm k},{\bm G \pm \bm \kappa_3}}\hat{V}\ket{\alpha,{\bm k},{\bm G}} &=& -V e^{\pm i C \frac{2\pi\alpha}{N}}\label{eq::matrix_element1}\\
\bra{\alpha \pm 1,{\bm k},{\bm G \pm \bm \kappa_1}}\hat{V}\ket{\alpha,{\bm k},{\bm G}} &=& -V \label{eq::matrix_element2}\\
\bra{\alpha \pm 1,{\bm k},{\bm G \pm \bm \kappa_2}}\hat{V}\ket{\alpha,{\bm k},{\bm G}} &=& -V e^{\pm iC \frac{(2\alpha\pm1)\pi}{N}}\label{eq::matrix_element3}
\end{eqnarray}
which show that, indeed, ${\bm k}$ is conserved due to Bloch's theorem in a cell of sides ${\bm a'}_1,{\bm a'}_2$. 

\begin{figure}[htb]
\includegraphics[width=0.99\columnwidth]{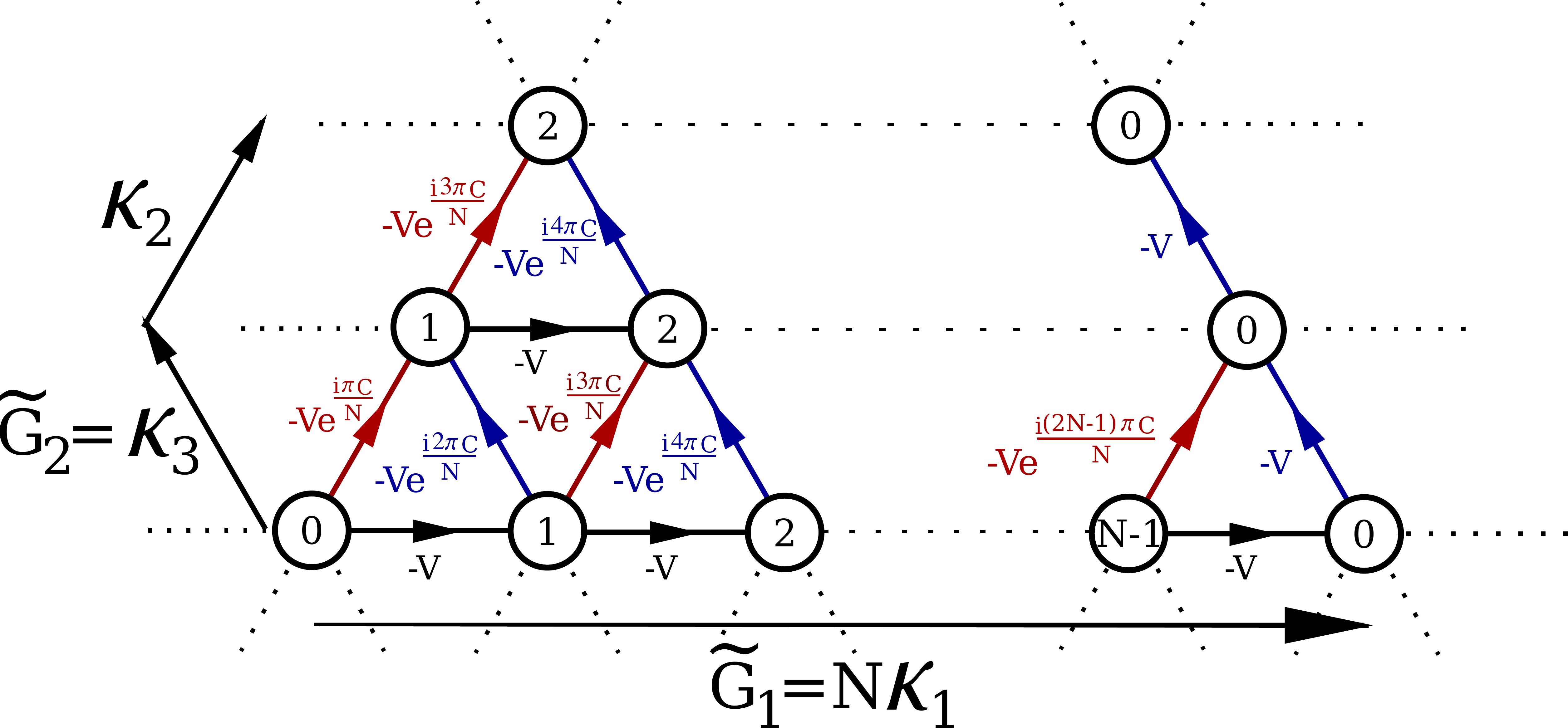}
\caption{Reciprocal space tight-binding representation of the OFL model. The reciprocal space is spanned by $\tilde{G}_1$ and $\tilde{G}_2$. A circle represents one internal state $\alpha=0,...,N-1$. Blue links between two internal states are associated to the hopping amplitude (in the reciprocal space) given the matrix element of by Eq.~\ref{eq::matrix_element1}. Similarly, black (resp. red) links are related to the matrix element of Eq.~\ref{eq::matrix_element2} (resp. Eq.~\ref{eq::matrix_element3}). Each triangle is pierced by a flux $\pi\frac{C}{N}$, leading to the lowest band Chern number $C$.}
\label{fiq:lattice_scheme}
\end{figure}

Beyond this translation invariance, many matrix elements are zero due to the form of the spin coupling under momentum exchange. For instance, Eq.~\ref{eq::matrix_element2} shows that an increase in the momentum by ${\bm \kappa_1}$ must be accompanied by a change of $\alpha$ by one. This is related to the presence of a higher degree of translational symmetry ({\it i.e.} a smaller real space unit cell) than one would expect from a reciprocal lattice of sides ${\bm \kappa_1}$ and ${\bm \kappa_3}$: by combining real space translation with spin rotation the effective reciprocal lattice can be chosen to have sides ${\bm \kappa_1}$ and $N{\bm \kappa_3}$. This structure can be readily seen by representing the matrix elements~(\ref{eq::matrix_element1})-(\ref{eq::matrix_element3}) in reciprocal space. They cause momentum transfers only by values $\pm{\bm \kappa}_{\alpha}$ so form a triangular lattice tight-binding model as depicted in Fig.~\ref{fiq:lattice_scheme}. The unit cell of this reciprocal-space model has sides $N{\bm \kappa}_1$ and ${\bm \kappa}_3$. Note that the phases of the matrix elements~(\ref{eq::matrix_element1})-(\ref{eq::matrix_element3}) imply that each triangle is pierced by a $\phi=\pi\frac{C}{N}$ flux. For a weak lattice limit, one can show that the Chern number of the lowest energy band is the total flux in the unit cell made of $2N$ triangles is equal to $\frac{1}{2\pi}*\phi*2N =C$\cite{PhysRevLett.109.215302}. A similar reciprocal space structure has been used to construct a generalization of the Landau level to Chern numbers greater than one~\cite{Wu-PhysRevLett.110.106802}.

Using this higher spin-translational symmetry, one can reduce the Hilbert space using another basis: $\ket{\alpha,{\bm k},{\bm \tilde{G}}}$ where $ {\bm \tilde{G}}$ is a vector of the reciprocal space spanned by ${\bm \tilde{G}}_1=N{\bm \kappa}_1$ and ${\bm \tilde{G}}_2={\bm \kappa}_3$. In real space the wave function is $\braket{{\bm r}}{\alpha,{\bm k},{\bm \tilde{G}}} = \ket{\alpha}e^{i {\bm r}\cdot ({\bm k + \alpha {\bm \kappa}_1 + {\bm \tilde{G}})}}$. The real space unit cell of this new reciprocal space is the parallelogram defined by ${\bm a}_1=\frac{2\pi}{N\kappa}(1,-1/\sqrt{3})$ and ${\bm a}_2=\frac{4\pi}{\kappa\sqrt{3}}(0,1)$, with an aspect ratio of $N$. 
The energy eigenstates can be decomposed on this basis 
\begin{equation}
\ket{\psi^{n{\bm k}}} = \sum_{\alpha,{\bm \tilde{G}}} c^{n {\bm k}}_{\alpha {\bm \tilde{G}}} \ket{\alpha,{\bm k},{\bm \tilde{G}}}
\end{equation}
where $n$ is the band index, $\alpha$ runs over the internal degree of freedom and ${\bm \tilde{G}}$ runs over all sites of the reciprocal lattice. For numerical calculations, we need to introduce a cut-off on momenta. Since ${\bm \tilde{G}}$ can be written as $n_x{\bm \tilde{G}}_1+ n_y {\bm \tilde{G}}_2$, where $n_x$ and $n_y$ are integers, we choose $n_x$ and $n_y$ between $-N_{\tilde{G}}/2$ and $N_{\tilde{G}}/2$ ($N_{\tilde{G}}$ even). Thus, the single-particle Hilbert space dimension for a given value of ${\bm k}$ is $N(N_{\tilde{G}}+1)^2$. In order to make sure that the value of $N_Q$ is large enough, we check that inversion symmetry is satisfied with a relative error on energies smaller that $10^{-12}$.

\begin{figure}[htb]
\includegraphics[width=0.8\columnwidth]{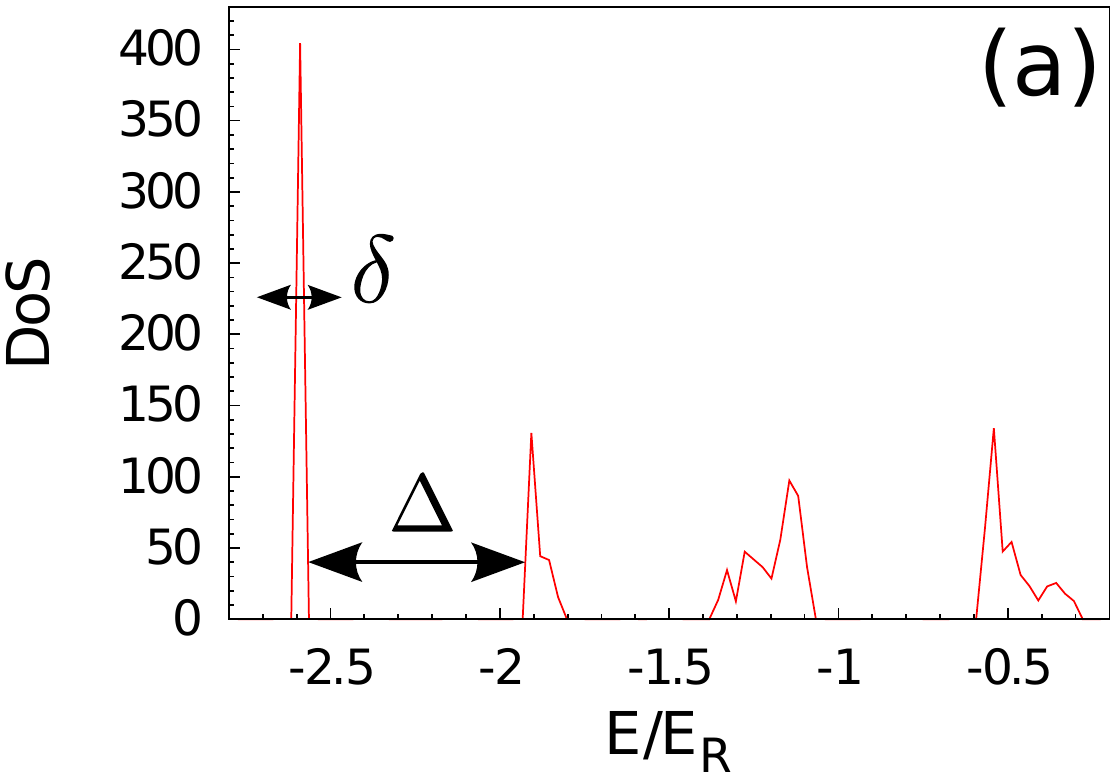}
\includegraphics[width=0.8\columnwidth]{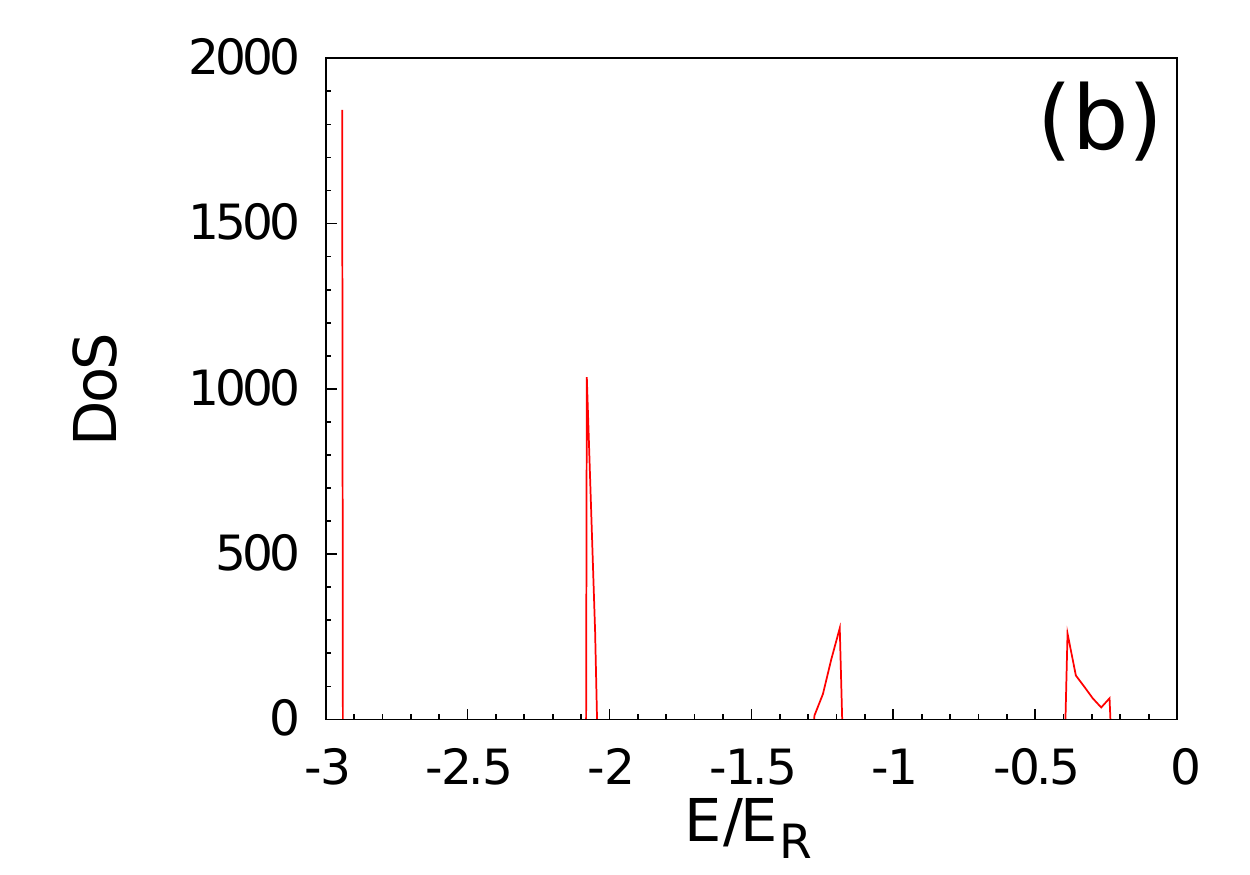}
\includegraphics[width=0.8\columnwidth]{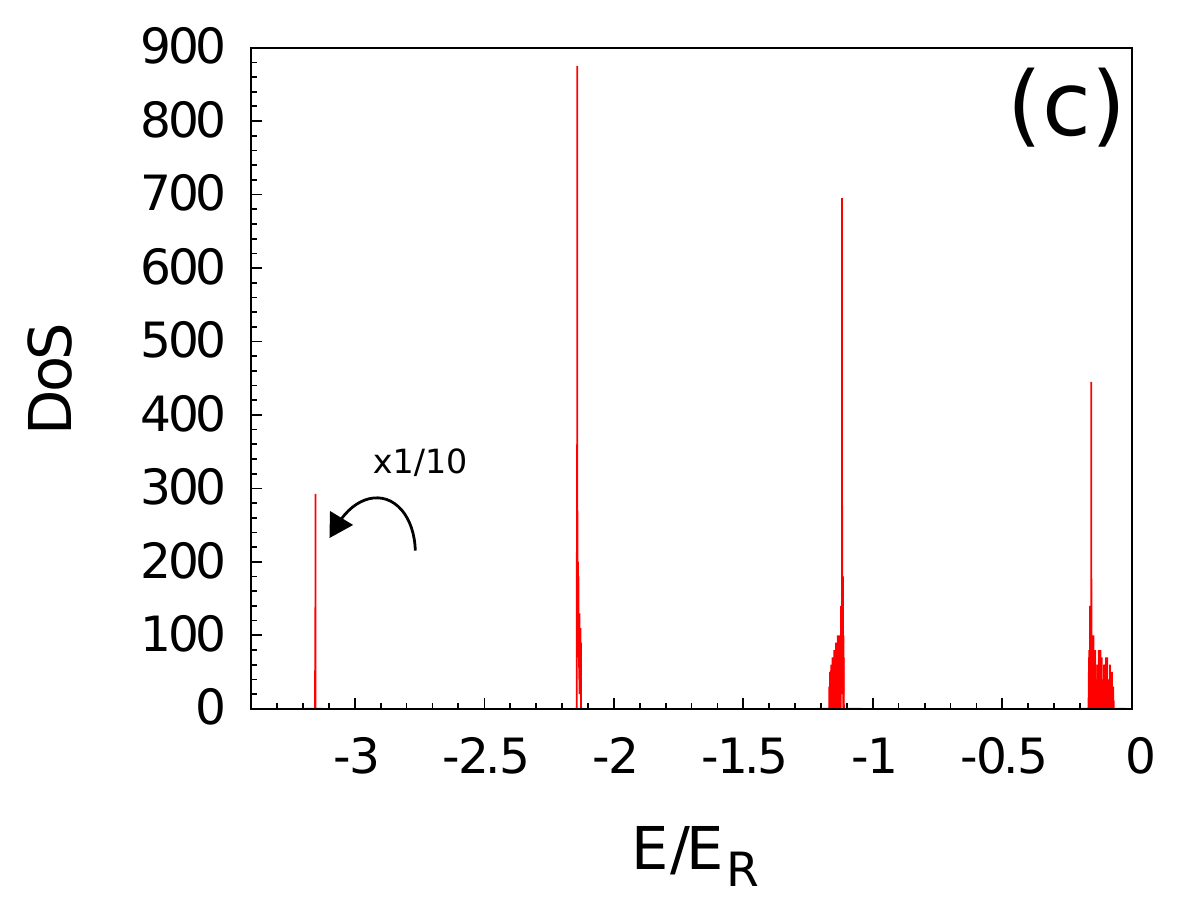}
\caption{Density of states for $C=1$, $V/E_R=1$ and $N=3$ (top panel), $N=4$ (center panel) and $N=5$ (lower panel). The band spread $\delta$ of the lowest band and the band gap $\Delta$ between the lowest band an the higher bands are depicted in the top panel. Their values are $\delta = 1.5\times 10^{-2} E_R$, $\Delta = 0.65 E_R$ for $N=3$, $\delta = 4.1\times 10^{-3} E_R$ and $\Delta = 0.86 E_R$ for $N=4$ and $\delta = 1.7\times 10^{-3} E_R$ and $\Delta = 1.0 E_R$ for $N=5$. Note that for $N=5$, we have rescaled the density of states of the lowest band by a factor $0.1$.}
\label{fig:onebody_c_1}
\end{figure}

\begin{figure}[htb]
\includegraphics[width=0.7\columnwidth]{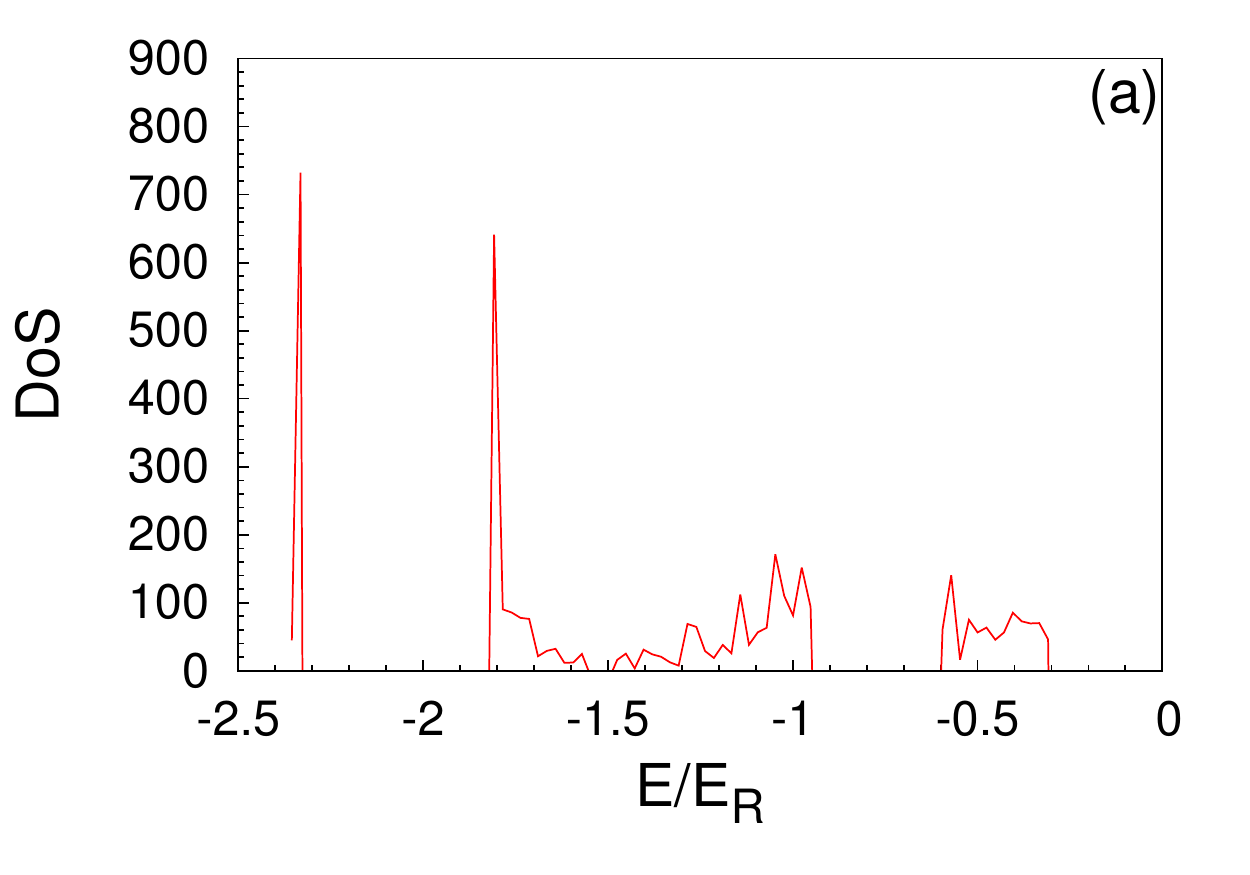}
\includegraphics[width=0.7\columnwidth]{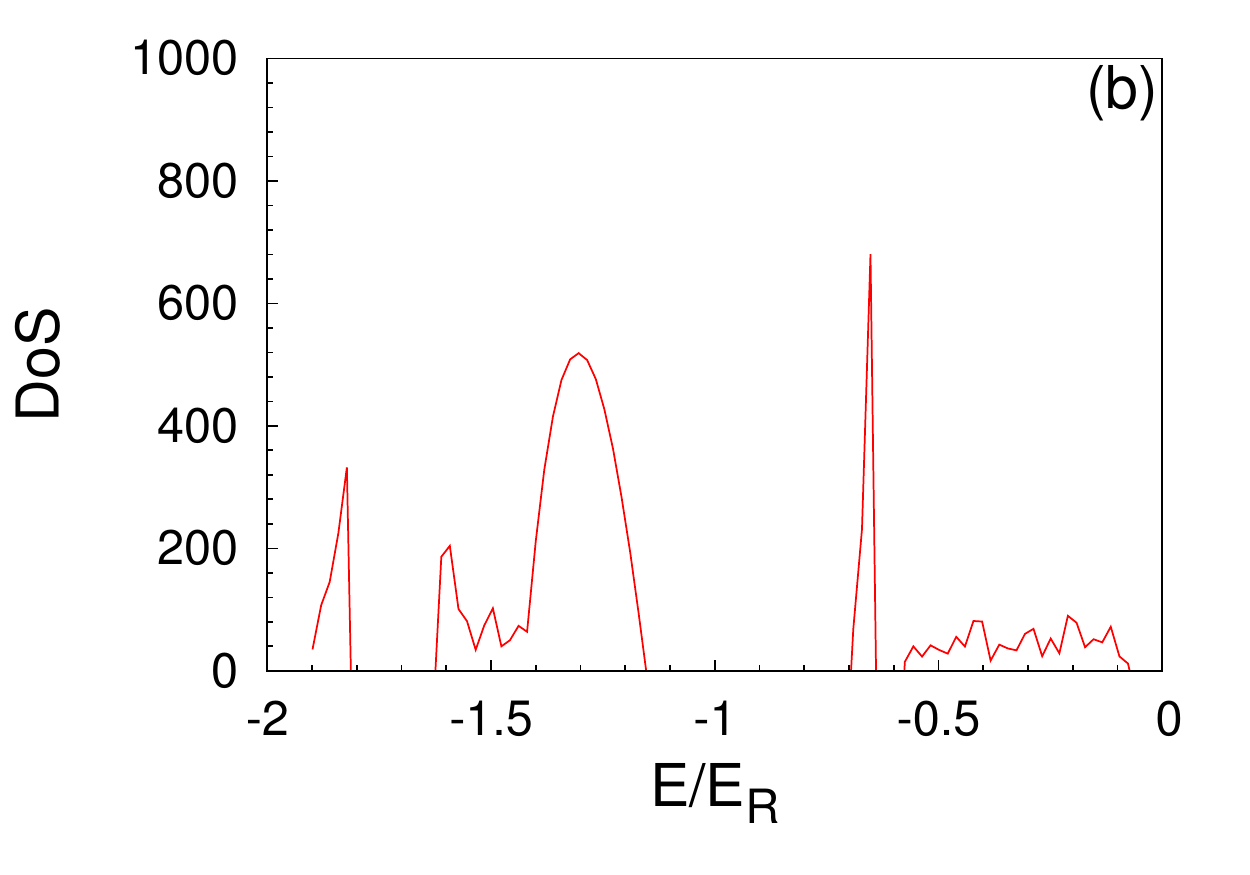}
\caption{{\it Upper panel:} Density of states for $C=2$ and $V/E_R=1$ for $N=5$. Here $\delta = 3.3 \times 10^{-2} E_R$ and $\Delta = 0.50 E_R$. {\it Lower panel:} Density of states for $C=3$ and $V/E_R=1$ for $N=5$. Here $\delta=8.6 \times 10^{-2} E_R$ $\Delta=0.19 E_R$.}
\label{fig:onebody_c_2_c_3}
\end{figure}

In Fig.~\ref{fig:onebody_c_1} we show the density of states for $C=1$ and different numbers of internal degrees of freedom. The ratio of the lowest band spread $\delta$ and the gap $\Delta$ between the lowest band and the higher band (depicted in Fig.~\ref{fig:onebody_c_1}a) varies from $\delta/\Delta = 2.3 \times 10^{-2}$ (for $N=3$) down to $\delta/\Delta = 1.7 \times 10^{-3}$ (for $N=5$). In these cases, assuming the lowest band is perfectly flat is a fairly good approximation for a wide range of interaction strengths. In Fig.~\ref{fig:onebody_c_2_c_3}, we give the density of states of this model for the Chern numbers $C=2$ and $C=3$ and $N=5$ for the internal degrees of freedom. While for $C=2$, the lowest band is still relatively flat, the band dispersion is more prominent for $C=3$. In the latter case, increasing $N$ would improve the flatness.

For the many-body calculations, we will consider a finite size system with periodic boundary conditions defined by the two vectors ${\bm L_x} = N_x {\bm a}_1$ and ${\bm L_y} = N_y {\bm a}_2$. The number of sites is $N_x \times N_y$. The aspect ratio of this finite size system is $r=\frac{N_y}{N_x} N$, since the unit cell itself has an aspect ratio of $N$. This extra multiplicative factor $N$ in $r$ does not appear in its usual definition for a FCI. Since a small aspect ratio spoils the signatures of the topological phases in finite size systems, we can use this additional knob to move away from these pathological cases.

\section{Fractional Chern Insulators}\label{sec:FCI}

In this section we give a brief overview of FCIs and summarize some of the main results that will be useful to understand the emergence of FQH-like phases in OFLs.

\subsection{FCI in a $C=1$ band}\label{sec:FCIC1}

FCIs are incompressible liquids formed in a partially filled band with a non-zero Chern number, which gives rise to a topological phase in the presence of interactions that are strong compared to the bandwidth. The emergence of such phases depends on several parameters such as the one-body dispersion relation, Berry curvature~\cite{PhysRevB.85.075116}, the range of the interaction or the band mixing. The flat-band procedure allows one to freeze the kinetic energy and to project onto the single partially filled band, effectively setting the one-body gap to infinity. This procedure is analogous to the Landau level projection. Starting from the one-body Hamiltonian written in the Bloch basis
\begin{equation}
{\cal H}_{\rm band}(\mathbf{k}) = \sum_{n\; {\rm bands}} E_n(\mathbf{k}) P_n(\mathbf{k})\label{eq:hamblochbasis}
\end{equation}
where $E_n(\mathbf{k})$ is the dispersion relation of the $n$-th band and $P_n(\mathbf{k})$ is the projector on the $n$-th band with momentum $\mathbf{k}$. Without affecting the topological properties of the system, we can consider the flattened Hamiltonian
\begin{equation}
{\cal H}^{\rm flat}_{\rm band}(\mathbf{k}) = \Delta_{\rm band} \sum_{n\; {\rm bands}} n P_n(\mathbf{k})\label{eq:flathamblochbasis}
\end{equation}
The energy separation $\Delta_{\rm band}$ between two consecutive bands can be set to infinity to consider a single band. In the following, we will focus on the lowest band i.e. $n=0$. In the presence of the interaction ${\cal H}_{\rm int}$ and in the flat band approximation, the effective Hamiltonian reads
\begin{equation}
{\cal H}_{\rm eff}={\cal P}_{0} {\cal H}_{\rm int} {\cal P}_{0}\label{eq:effectiveham}
\end{equation}
where ${\cal P}_{0}=\sum_{\mathbf{k}} P_0(\mathbf{k})$ is the projector onto the lowest band. This expression is similar to the effective Hamiltonian projected onto the lowest Landau level for the FQHE. Note that the natural energy scale for this effective Hamiltonian is the two particle energy scale. Such a choice will be discussed in Sec.~\ref{sec:laughlin}.

Among the signatures that reveal the emergence of a FQH-like phase in FCIs, the simplest ones are those that can be extracted from the spectral analysis of finite size exact diagonalizations. We consider a system of $N_{\rm b}$ bosons on a lattice of $N_x \times N_y$ unit cells with periodic boundary conditions in both $x$ and $y$ directions. The filling factor is defined as 
\begin{equation}
\nu = \frac{N_{\rm b}}{N_x \times N_y}\label{eq:fillingfactor}
\end{equation}
$N_x \times N_y$ is equivalent to the number of flux quanta for FQH systems.

For the FQHE with periodic boundary conditions in both directions (the torus geometry), the physics of topological phases leads to a low-energy manifold with a characteristic degeneracy. At filling factor $\nu=1/2$ for bosons where the Laughlin state is realized, we observe two degenerate states separated by a gap from the neutral excitations. This exact degeneracy is a consequence of the magnetic translation symmetries~\cite{PhysRevLett.55.2095,Bernevig-2012PhysRevB.85.075128}. Such a symmetry is absent in FCIs due to the non-flatness of the Berry curvature~\cite{Parameswaran-2012PhRvB..85x1308P,goerbig-2012epjb,Bernevig-2012PhysRevB.85.075128}. Thus for these systems, the degeneracy is generically lifted in a finite-size system but we still expect to observe a low energy manifold separated by a gap from higher energy excitations. 

The states in the low energy manifold of a given topological phase have well-defined quantum numbers, in particular momentum. The number of states per momentum sector of many FQH model wavefunctions can be derived from a generalized Pauli principle~\cite{Haldane-PhysRevLett.67.937}. From this knowledge and using the FQH to FCI mapping developed in Ref.~\onlinecite{Bernevig-2012PhysRevB.85.075128}, one can predict what should be the number of states per momentum sector for each FQH phase realized on a FCI. Such a characterization can be done both at the exact filling factor where the state is expected and when deviating from this filling factor by inserting quasihole or quasielecton excitations. Nucleation of these excitations can be done in the FCIs by increasing or reducing the number of lattice unit cells at fixed particle number, which is the equivalent of adding or removing flux quanta in the FQHE language.

\begin{figure}[htb]
\includegraphics[width=0.8\columnwidth]{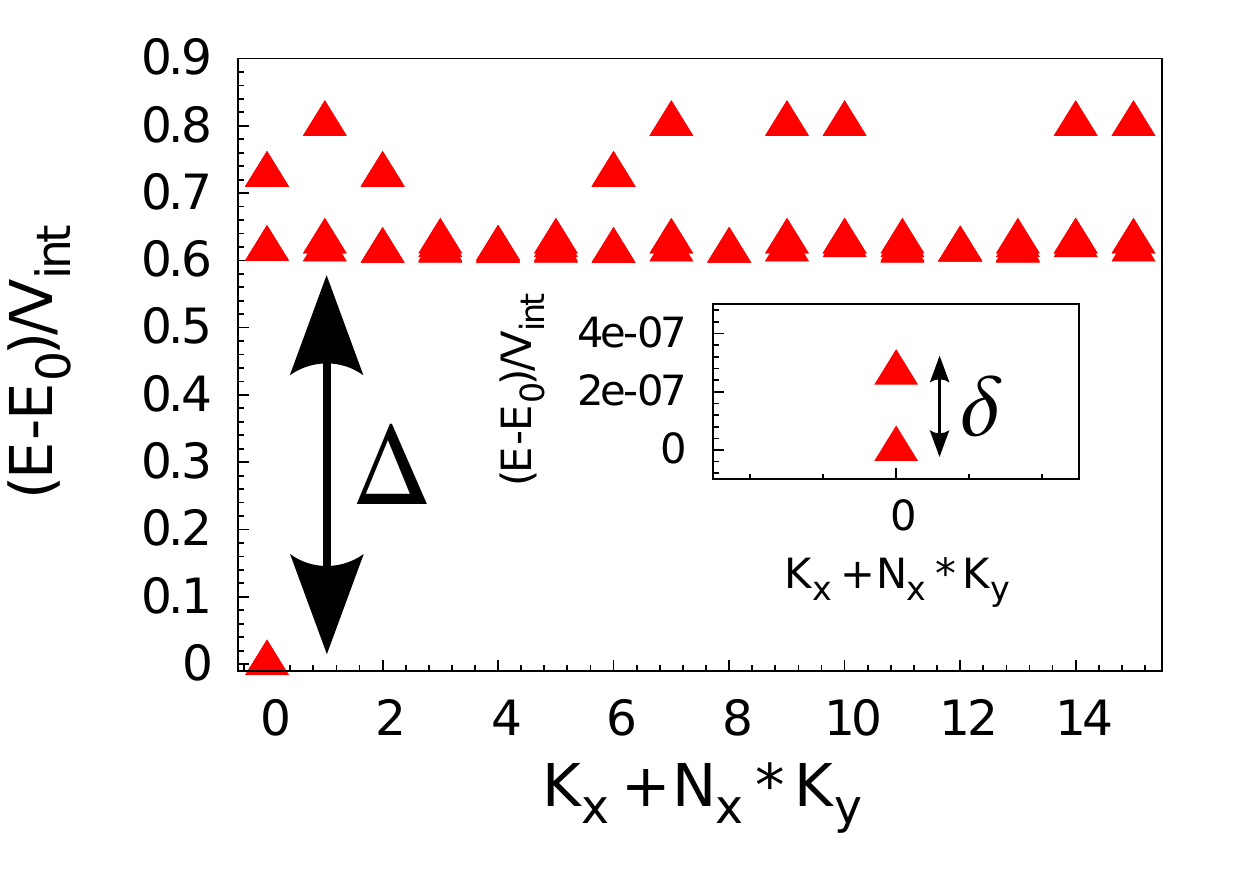}
\caption{Low energy spectrum for the OFL at filling factor $\nu=1/2$ with $N_{\rm b}=8$ bosons on a $N_x=8$ and $N_y=2$ lattice with $N=4$. The calculations are performed in the flat band limit. The energies are displayed as a function of the linearized total momentum $K_x+N_x \times K_y$. They are shifted by $E_0$, the system lowest energy. The two low energy states in the $(K_x,K_y)=(0,0)$ momentum sector are almost degenerate. The inset provides a zoom on these two states, the energy splitting $\delta$ being $\simeq 2.6 \times 10^{-7} V_{\rm int}$. The energy gap $\Delta$ is defined as the difference between the third lowest energy state and the second lowest energy, irrespective of the momentum sector.}
\label{fig:laughlinofl}
\end{figure}

A typical energy spectrum for a FCI at filling factor $\nu=1/2$ is shown in Fig.~\ref{fig:laughlinofl}. The energies are displayed as a function of the linearized momentum $K_x + N_x K_y$ where $K_x$ (resp. $K_y$) is the total momentum in the $x$ (resp. $y$) direction. 

Beyond the flat band limit, the band mixing or the one-body dispersion relation can affect the stability of the FCI phase. As was pointed out in the context of the FQH on a lattice~\cite{Hafezi-PhysRevA.76.023613,Sterdyniak-PhysRevB.86.165314} and more recently in FCI systems~\cite{Kourtis-PhysRevLett.112.126806}, the Laughlin phase survives even in the presence of strong interactions that induce a large band mixing, independent on whether the higher bands have an identical or an opposite Chern number. More complicated states such as the Moore-Read are more sensitive to the band mixing~\cite{Sterdyniak-PhysRevB.86.165314}. 

\subsection{FCI in a $C>1$ band}
\label{sec::fci_higherC}
\begin{figure}[htb]
\includegraphics[width=0.8\columnwidth]{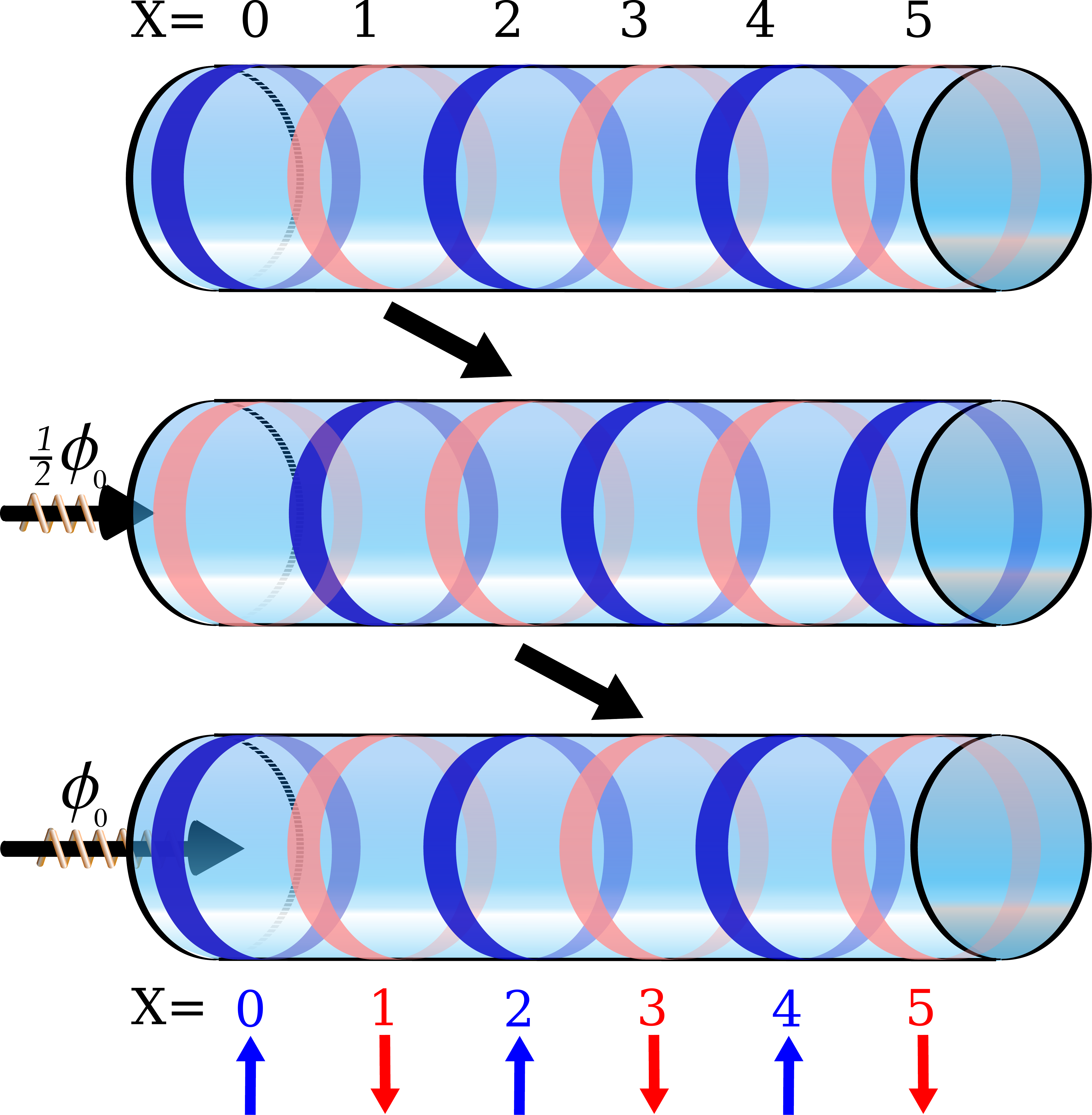}
\caption{Action of the flux insertion on the Wannier orbitals in a $C=2$ Chern insulator. {\it Top panel:} We consider Wannier states localized in the $x$ direction (cylinder axis). The family of Wannier states for a given $k_y$ are localized along the $x$ direction in each unit cell $X=0,1,2,...$. {\it Center panel:} When adiabatically inserting half a flux along the cylinder axis each Wannier state will be displaced by one step i.e. the one localized around $X=1$ will now be centered around $X=2$. {\it Lower panel:} Upon a complete flux insertion, we have transported each Wannier state from $X$ to $X+2$. Labeling spin up (resp. spin down) the Wannier states centered around even (resp. odd) values of $X$, we observe than the up and down states act as two separate copies of a $C=1$ Chern insulator.}
\label{fig:fluxinsertionc2}
\end{figure}

An interesting feature of the FCI systems is the ability to consider a partially filled band with a Chern number larger than one. Several studies have numerically investigated this problem~\cite{Wang-PhysRevB.86.201101,Yang-PhysRevB.86.241112,Liu-PhysRevLett.109.186805,PhysRevB.87.205137,Wu-PhysRevLett.110.106802,Wu-2013arXiv1309.1698W}. In the non-interacting case, Barkeshli and Qi~\cite{Barkeshli-PhysRevX.2.031013} mapped a $C > 1$ Chern band onto a $C$-component lowest Landau level (LLL) using hybrid Wannier states\cite{Qi-PhysRevLett.107.126803}. From that perspective, it is natural to suggest that strong interactions would lead to spinful (or more generally multi-component) FQHE. Fig.~\ref{fig:fluxinsertionc2} depicts how the decoupling of the $C=2$ case into a $2$-component CI works. Let us consider a family of Wannier states with a given momentum $k_y$ along the $y$ direction. Each of these states is localized in the $x$ direction around positions $X$ belonging to different unit cells. Under an adiabatic flux insertion along the cylinder axis, one shifts these Wannier states from $X$ to $X+C$. Thus one can disentangle $C$ decoupled copies of a CI with unit Chern number for which one associates a fictitious degree of freedom. In this example, we associate a spin up (resp. spin down) to the states localized around even (resp. odd) values of $X$. With this picture in mind, we clearly see that turning on strong interaction in such a system should lead to multi-component FQH phases.

It can already be seen at the non-interacting level that the situation for finite size systems might not be as simple as depicted in the paragraph above. For example if the number of unit cells $N_x$ in the $x$ direction is not a multiple of the Chern number $C$, the separation into $C$ copies of a CI with a unit Chern number breaks down. Indeed, a flux insertion in a finite system with periodic boundary conditions will mix the different fictitious degrees of freedom. Let us consider the example of Fig.~\ref{fig:fluxinsertionc2} for $C=2$ where we would choose $N_x=5$. Under a flux insertion, the orbital localized around $X=4$ (with an up spin) would flow to a position localized around $X=1$ (with a down spin). Thus on a finite system with periodic boundary conditions, the fictitious degree of freedom and the translations projected onto the non-trivial band are actually entangled~\cite{Wu-PhysRevLett.110.106802}, giving rise to new topological states unknown in the FQH picture.

In the FQHE with a $SU(C)$ internal degree of freedom, the Halperin state\cite{Halperin83} is the natural generalization of the Laughlin state. In the simplest case, it reads
\begin{equation}
 \Psi_{[m;n]}^{SU(C)} = \Phi_{\{m\}}^{\mathrm{intra}} \Phi_{\{n\}}^{\mathrm{inter}}\exp\left(-\frac{1}{4}\sum_{i=1}^C\sum_{k_i = 0}^{N_i} |z_{k_i}^{(i)}|^2 \right)\label{halperin21}
\end{equation}
where 
\begin{equation}
\Phi_{\{m\}}^{\mathrm{intra}} = \prod_{i=1}^C\prod_{k_i<l_i} (z_{k_i}^{(i)}-z_{k_i}^{(i)})^{m}\label{halperin_intra}
\end{equation}
is the product of a Laughlin state for each component and 
\begin{equation}
\Phi_{\{n\}}^{\mathrm{inter}} = \prod_{i<j}^C\prod_{k_i = 1}^{N_i}\prod_{k_j = 1}^{N_j}(z_{k_i}^{(i)}-z_{k_j}^{(j)})^{n}\label{halperin_inter}
\end{equation}
accounts for correlations between components. Here, $z_{k}^{(i)}$ is the complex position (in the plane) of the $k$-th particle of component $i$. In the following, we will focus on the $m=2$ and $n=1$ case. The $[2;1]$ Halperin state is the exact densest zero energy of the contact interaction (the Laughlin $\nu=1/2$ state corresponding to the particular case $C=1$). It is a $SU(C)$ singlet and describe a state at filling factor $\nu_{\rm FQH}=\frac{C}{C+1}$. On the torus geometry, this $[2;1]$ Halperin state is $(C+1)$-fold degenerate. 

States analogous to $[2;1]$ Halperin states have been observed in several FCI models with $C>1$~\cite{Wang-PhysRevB.86.201101,Yang-PhysRevB.86.241112,Liu-PhysRevLett.109.186805,PhysRevB.87.205137}. Note that the definition of the filling factor is slightly different for the FCI and the FQHE. For the FCI, it is defined with respect to the total number of one-body states of the fully occupied band each carrying a ``flux'' $C$. For the FQHE, using the notation $\nu$ for the filling factor of the FCI (irrespective of the value of $C$) and $\nu_{\rm FQH}$ for the FQHE, we have 
\begin{equation}
\nu=\frac{\nu_{\rm FQH}}{C}\,. \label{eq:fillingfactorFCIFQH}
\end{equation}
The $[2;1]$ Halperin state is a natural candidate to look for in a bosonic $C>1$ FCI. Unless engineered in a very specific way~\cite{Wu-2013arXiv1309.1698W}, the interactions that are considered for FCIs are not sensitive to the fictitious degree of freedom that is introduced to separate the CI into $C$ copies of a CI with a unit Chern number. Thus an on-site interaction for a FCI is analogous to the model contact interaction for the $[2;1]$ Halperin state. There are still subtle differences between the state emerging in these FCIs and the Halperin state. The latter is only defined when the number of particles is a multiple of $C$. While in the FCI, an almost $C+1$ degenerate low energy manifold appears at filling $\nu=1/(C+1)$ irrespective of the particle number. This is a consequence~\cite{Wu-PhysRevLett.110.106802} of the entanglement between the fictitious degree of freedom and the translations. Other differences can also be unveiled through the entanglement spectrum that we discuss later.

\subsection{FCI and entanglement spectrum}

In order to probe the topological order within the numerical simulations, the entanglement spectrum~\cite{li-PhysRevLett.101.010504}(ES) is a valuable tool. Among the different entanglement spectra, the particle entanglement spectrum~\cite{sterdyniak-PhysRevLett.106.100405} (PES) allows one to obtain the information encoded in the groundstate wavefunction related to the system's bulk excitations. In the context of the FCI, it has been shown~\cite{Bernevig-2012arXiv1204.5682B} that the ES differentiates between a charge density wave and a Laughlin state.

\begin{figure}[htb]
\includegraphics[width=0.8\columnwidth]{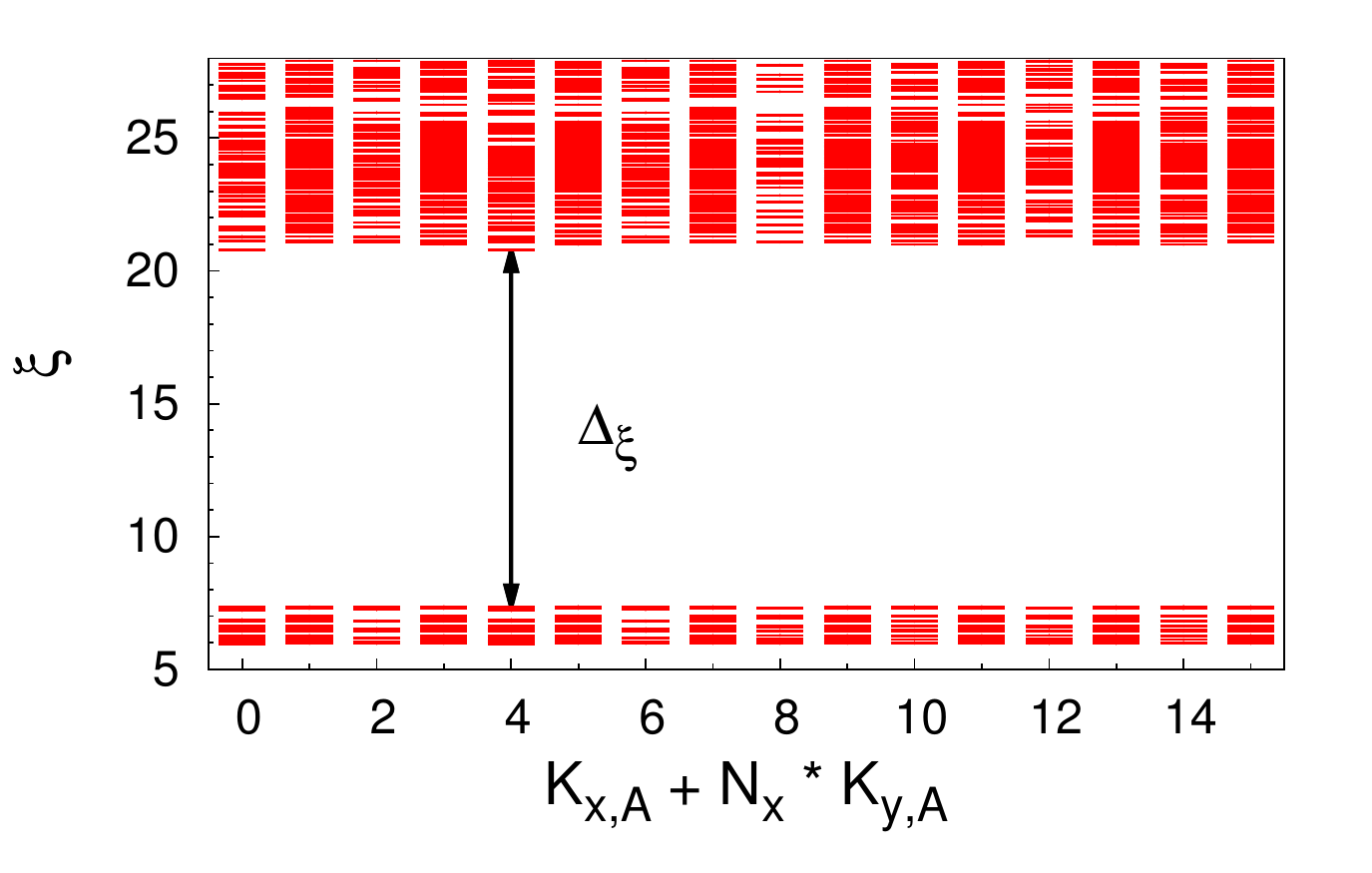}
\caption{PES computed for the low energy state manifold at $\nu=1/2$ with $N_{\rm b}=8$ bosons on a $N_x=8$ and $N_y=2$ lattice with $N=4$ shown in Fig.~\ref{fig:laughlinofl}. We keep $N_A=4$ bosons and trace out the remaining particles. The entanglement energies $\xi$ are displayed as a function of the linearized total momentum $K_{x,A}+N_x \times N_{y,A}$. We observe a clear entanglement gap $\Delta_\xi$. The low energy structure below the entanglement gap corresponds to the Laughlin state $\nu=1/2$.}
\label{fig:laughlinoflPES}
\end{figure}

For a $d$-fold degenerate state $\{|\psi_i\rangle\}$, we consider the density matrix
\begin{equation}
\rho=\frac{1}{d}\sum_{i=1}^{d}|\psi_i\rangle\langle\psi_i|\label{eq:densitymatrix}
\end{equation}
We divide the $N$ particles into two groups $A$ and $B$ with respectively $N_A$ and $N_B$ particles. Tracing out on the particles that belong to $B$, we compute the reduced density matrix $\rho_A={\rm Tr}_B \rho$. This operation preserves the geometrical symmetries of the original state, so we can label the eigenvalues $\exp(-\xi)$ of $\rho_A$ by their corresponding momenta, $K_{x,A}$ and $K_{y,A}$. A typical PES is shown in Fig.~\ref{fig:laughlinoflPES}, where the $\xi$'s (generally called entanglement energies) are plotted as a function of the linearized momentum.

For FQH model states, the number of non zero eigenvalues of $\rho_A$ matches the number of quasihole states for $N_A$ particles and the same number of flux quanta as the original state. The quasihole counting is characteristic of each topological state, and the PES acts as a fingerprint of the phase. In the FCI, one expects to observe a low entanglement energy structure similar to the one of the model state (with the same number of levels in each momentum sector) with a gap $\Delta_\xi$ to higher energy excitations (see Fig.~\ref{fig:laughlinoflPES}). Such a feature has been shown for Laughlin and MR-like states in FCI~\cite{regnault-PhysRevX.1.021014, PhysRevB.85.075116}. Tracking the entanglement gap when tuning any external parameters allows one to determine whether or not the ground state manifold of the system is still correctly described by a given model state. While overlaps between a FCI state and a FQH model states can be computed~\cite{Wu-PhysRevB.86.085129}, or adiabatic continuation from FQH to FCI can be performed~\cite{Wu-PhysRevB.86.085129,Scaffidi-PhysRevLett.109.246805,Wu-PhysRevB.86.165129}, the PES is easier to implement and does not depend on gauge fixing. For those reasons, we will use the PES over other approaches.

For FCIs in a higher Chern number band, the PES allows one to unveil the multi-component nature of the FCI phases and their difference from usual multi-component Halperin states. The total number of quasihole states for the Halperin $[2,1]$ state and a given system size is identical to the number of quasihole states for the Laughlin state at $\nu=\frac{1}{C + 1}$. Since these two model states have the same filling factor and the internal degree of freedom is not accessible in a $C>1$ FCI, one would naively expect that the PES would not be able to discriminate between them. However, this is untrue. As was shown in Ref.~\onlinecite{PhysRevB.87.205137}, the interplay between the $SU(C)$ symmetry of the Halperin state and the number $N_A$ of kept particles can actually reduce the number of quasihole states that appear in the PES from the pure Laughlin case: More precisely, such differences appear for $N_A > N/C$. For $C>1$ FCIs, this signature of the Halperin state has been clearly observed in the PES~\cite{PhysRevB.87.205137,Liu-PhysRevLett.109.186805}. This is an evidence for an internal degree of freedom in these systems. Moreover the differences introduced by the coupling between the fictitious degree of freedom in $C>1$~\cite{PhysRevB.87.205137,Wu-PhysRevB.89.155113} FCIs and the translations in a finite-size system have signatures in the PES~\cite{PhysRevB.87.205137} that can be understood using a modified version of the generalized Pauli principle~\cite{Wu-PhysRevB.89.155113}.

\section{Numerical Results for $C=1$}\label{sec:numC1}

In this section, we provide an in-depth study of two FQH phases that emerge in the OFL model given by Eq.~\ref{eq::OFL_Ham} for Chern number one.

\subsection{Two-body spectrum}

Unlike most Chern insulators that are defined by a tight-binding model, optical flux lattice models are continuous models in real space. Thus, the interaction between the atoms mainly depends on the considered atomic species. Here, we focus on the simplest interaction: the $s$-wave scattering that correctly describes cold gases of alkali atoms like $^{87}$Rb. Thus, the interaction potential is given by 
\begin{equation}
{\cal H}_{\rm int} = V_{\rm int}\delta(r-r') \,.
\label{eq::int_delta}
\end{equation}

In the usual quantum Hall problem, two-body interactions can be described by Haldane pseudopotentials~\cite{PhysRevLett.51.605}. This is a set of parameters $V_p$ that weight the two-particle states with relative angular momentum $p$. On the two-particle spectrum, each non-zero pseudopotential leads to two almost degenerate bands with non-zero energies independent of momentum, once mapped onto the FCI Brillouin zone~\cite{Bernevig-2012PhysRevB.85.075128} while other states have zero energy. The $\nu=\frac{1}{m}$ Laughlin states are the exact densest zero-energy states of the Hamiltonian given by $V_p = 1$ for $p < m $, $V_p = 0$ otherwise. There are $m$ degenerate Laughlin states in the torus geometry.

While the absence of continuous translation and rotation symmetries in FCIs lead to an absence of a clear definition of pseudopotentials~\cite{PhysRevB.88.035101}, it was suggested to look at the two-body spectra as a tool to diagnose whether a Chern insulator model could host a fractional topological state~\cite{PhysRevLett.111.126802}. In particular, it was conjectured that pairs of bands separated by a gap could lead to a stable topological state.

\begin{figure}[htb]
\includegraphics[width=0.9\columnwidth]{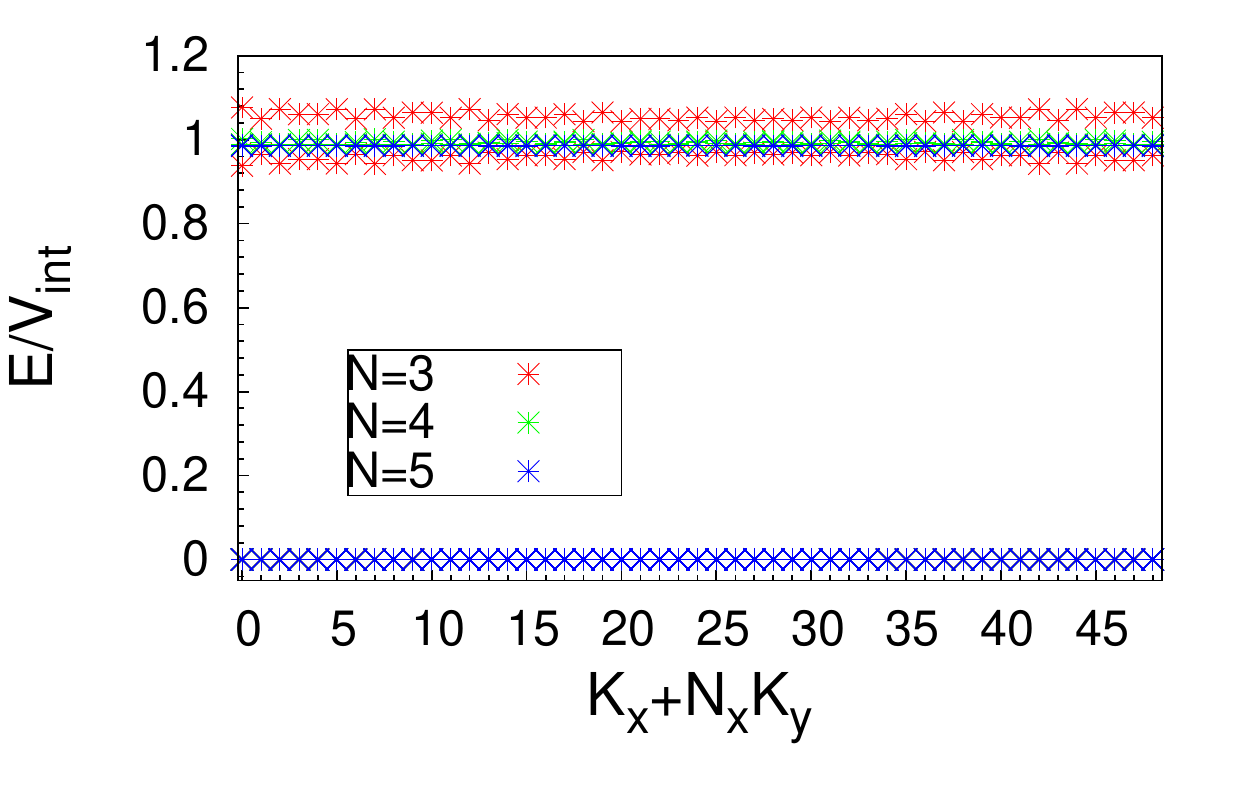}
\caption{Two-body spectrum of the interaction given by Eq.~\ref{eq::int_delta} in the OFL model for $C=1$ and $V/E_R=3$.}
\label{fig::delta_2body}
\end{figure}

Once projected onto the lowest band of the non-interacting model, the two-body interaction, given by Eq.~\ref{eq::int_delta} is very similar to the same interaction projected on the lowest Landau level on the torus. This can be seen at the level of the two particle spectra, shown in Fig.~\ref{fig::delta_2body}: it is made of two branches, whose energies are almost independent of momentum and very close to $V_{\rm int}$ and get closer as $N$ is increased. For FQH systems on a torus, all the other eigenstates have strictly zero energy. In our model (and more generally in FCIs), we can observe that some other eigenstates also have a non-zero energy, but several orders of magnitude smaller than those around $V_{\rm int}$ (for $N=4$, these non-zero energies are at most $10^{-4} V_{\rm int}$). The separation observed in this model implies the possibility of a robust Laughlin state at $\nu=\frac{1}{2}$ in this model.

\subsection{$\nu = 1/2$: the Laughlin state}\label{sec:laughlin}

\begin{figure}[htb]
\includegraphics[width=0.8\columnwidth]{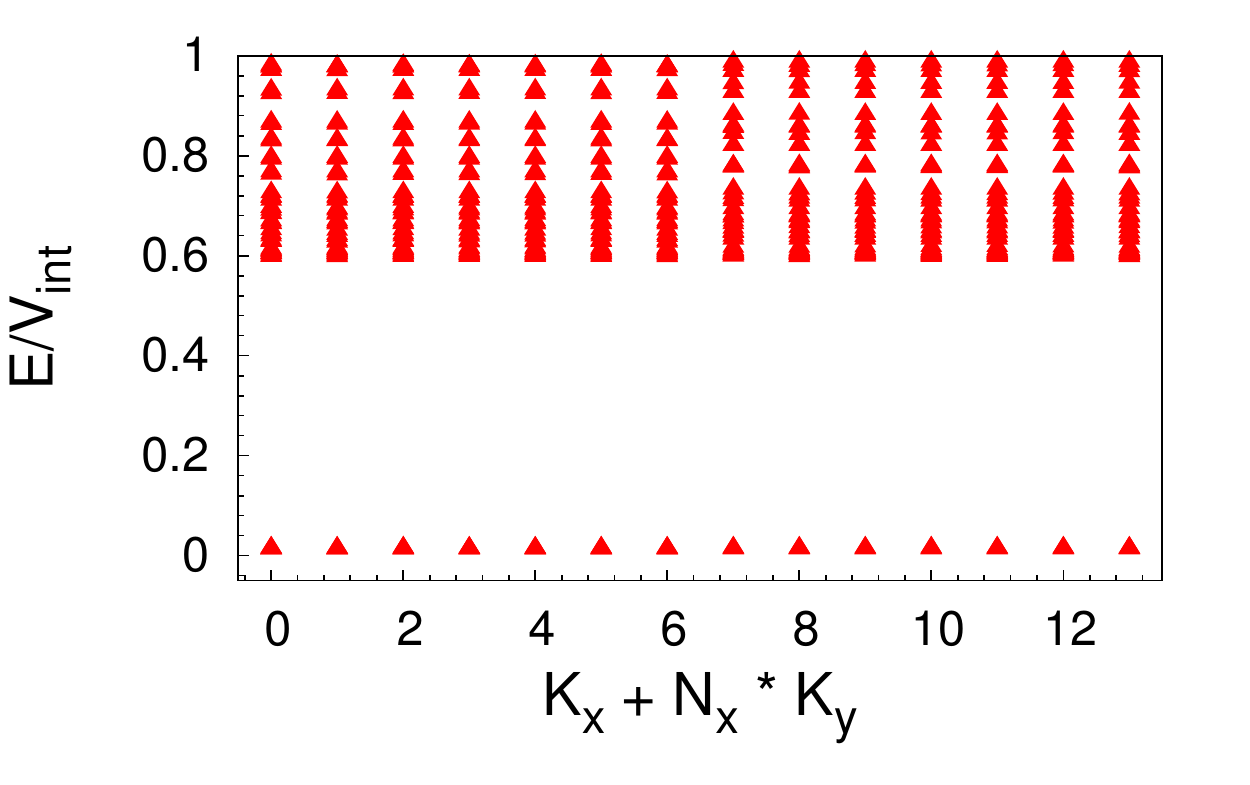}
\caption{Energy spectrum for $C=1$, $N=4$, $N_{\rm b}=6$, $N_x=7$, $N_y=2$ and $V=E_R$. There are 2 additional sites compared to the ground states for $N_{\rm b}=6$. The counting of the low energy part match the number of quasiholes states for Laughlin states with $N_{\phi}=14$ on the torus. The splitting in energy between the quasiholes states is $\delta = 0.002 V_{\rm int}$. The gap between quasiholes states and the higher energy states is $0.58V_{\rm int}$.}
\label{fig::laughlin_quasi}
\end{figure}

We start our study with the Laughlin state at $\nu=1/2$. This state, characterized by a two-fold (quasi-)degenerate groundstate, is obtained quite generically in this OFL model. We will first consider the flat band approximation discussed in Sec.~\ref{sec:FCIC1}. An example of energy and entanglement spectrum is shown in Figs.~\ref{fig:laughlinofl} and \ref{fig:laughlinoflPES}, respectively: the two lowest energy states appear in the momentum sectors expected from the FQH to FCI mapping~\cite{Bernevig-2012PhysRevB.85.075128} as explained in section~\ref{sec:FCI}. We check the nature of these states by computing their PES countings. In every case, we obtain the same results as for the Laughlin state on the torus except from some spurious entanglement eigenvalues of very small probability (very large entanglement gaps). By adding sites, which corresponds in the FQH language to adding flux quanta, we can nucleate quasihole excitations. The numbers of quasihole states per momentum sector are those predicted for the Laughlin phase. An example of energy spectrum with two quasiholes is given in Fig.~\ref{fig::laughlin_quasi}. 

\begin{figure}[htb]
\includegraphics[width=0.99\columnwidth]{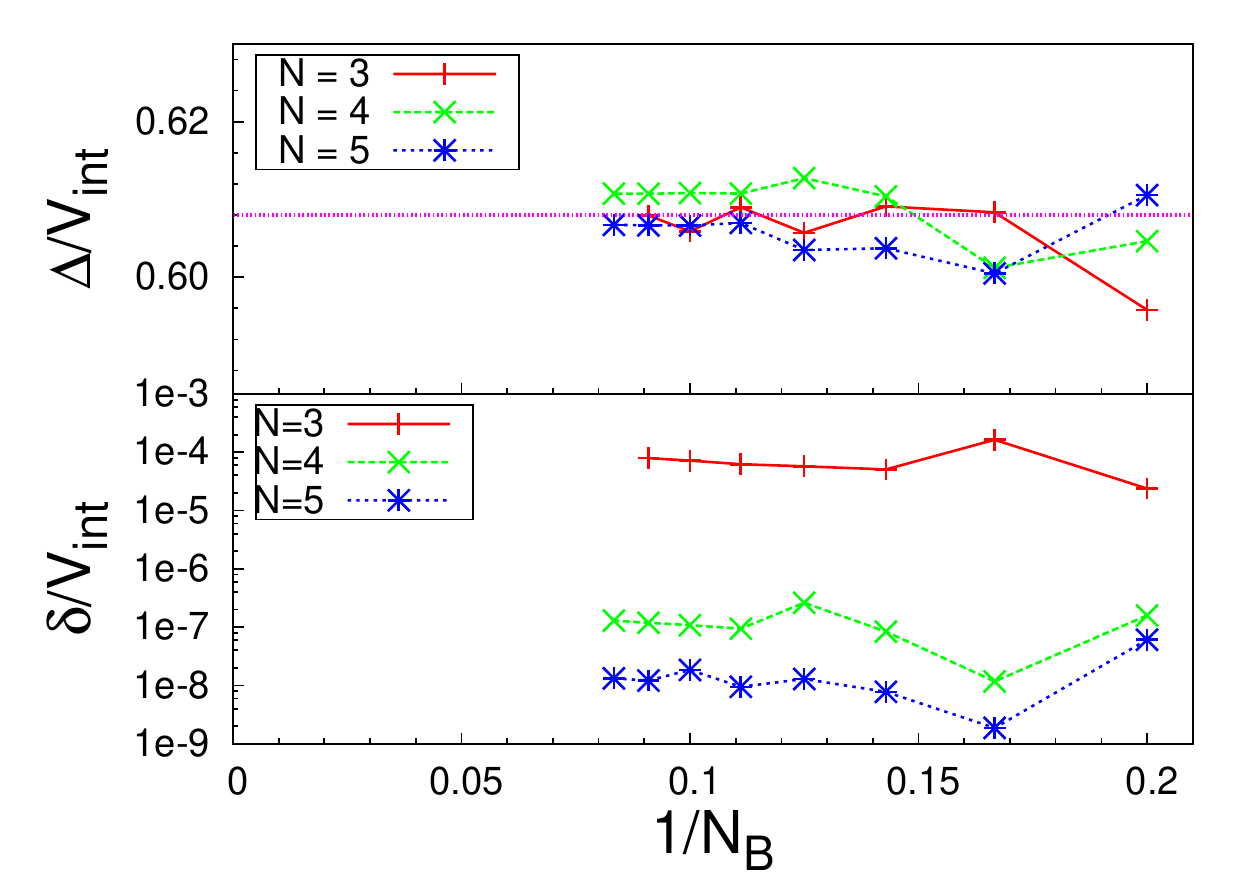}
\caption{Gap $\Delta$ (Upper panel) and energy splitting $\delta$ (lower panel) as a function of the particle number inverse for different $N$ values with $V=3E_R$, $N_x=N_{\rm b}$ and $N_y=2$ . The purple line shows the gap extrapolated for FQH on torus in Ref.~\onlinecite{Repellin-PhysRevB.90.045114}. The gap barely depends on the number of particles and on the number of spin species.}
\label{fig::laughlin_thermo}
\end{figure}

While similar results were already obtained on other Chern insulator models~\cite{PhysRevLett.107.146803}, this model gives a particularly strong Laughlin phase. Indeed, the gap separating the groundstate manifold and the excited states is almost independent of the number of particles and of the number of spin species, as can be seen in Fig.~\ref{fig::laughlin_thermo}a. Thus, we can extrapolate easily its values to be $\Delta=0.608 V_{\rm int}$ in the thermodynamic limit of the flat band model. This value is very close to the one found for the Laughlin state ($\Delta=0.615(5) V_{\rm int}$) on the torus~\cite{Repellin-PhysRevB.90.045114}, or the sphere geometry~\cite{PhysRevLett.91.030402,cooper2008rapidly}. Similarly, the energy splitting between the two groundstates decreases with the number of spin species and in each of the cases we studied (i.e. $N=3,4,5$), is much smaller than what was reported previously in tight-binding lattice-based FCIs~\cite{PhysRevB.85.075116, PhysRevB.88.115117}. Moreover, the manifold of quasiholes states is also almost degenerate and is separated from excited states by an almost constant gap as can be seen in Fig.~\ref{fig::laughlin_quasi}.

\begin{figure}[htb]
 \includegraphics[width=0.99\columnwidth]{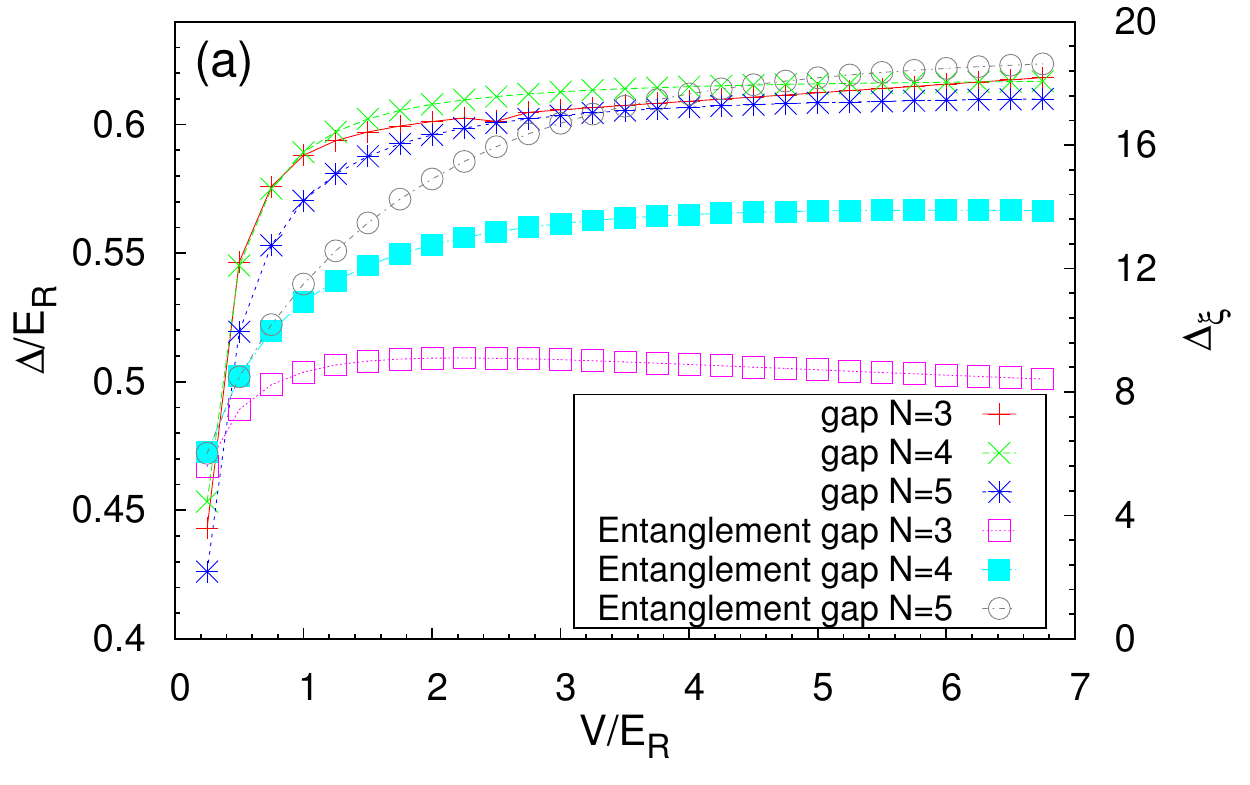}
 \includegraphics[width=0.99\columnwidth]{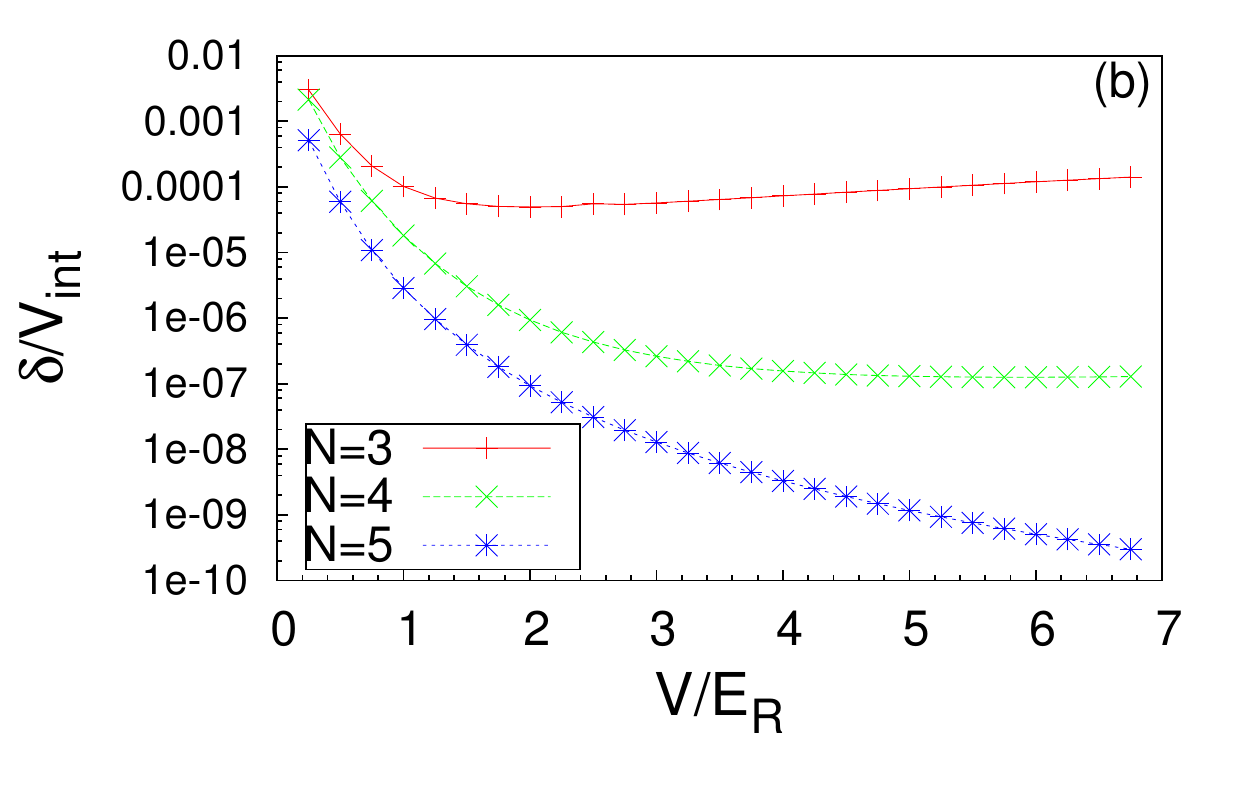}
 \caption{\textit{Upper Panel:} Gap $\Delta$ and entanglement gap $\Delta_\xi$ as a function of the laser strength for different spin values for $N_{\rm b}=8$, $N_x= 8$ and $N_y= 2$. \textit{Down Panel:} Energy splitting $\delta$ between the two groundstates as a function of the laser strength $V$ for different spin values for $N_{\rm b}=8$, $N_x= 8$ and $N_y= 2$. }
\label{fig::laughlin_laser} 
\end{figure}

We then study the stability of the Laughlin state in this model. First, we look at the effect of the laser strength $V$. As can be seen in Fig.~\ref{fig::laughlin_laser} for $N_{\rm b}=8$ particles, both the energy and the entanglement gaps (Fig.~\ref{fig::laughlin_laser}a) grow with the laser strength at small values and, then, saturate. The energy splitting decreases for every $N$ except for $N=3$ where it increases slightly beyond $V\sim 1.2 E_R$ (Fig.~\ref{fig::laughlin_laser}b) .

\begin{figure}[htb]
 \includegraphics[width=0.99\columnwidth]{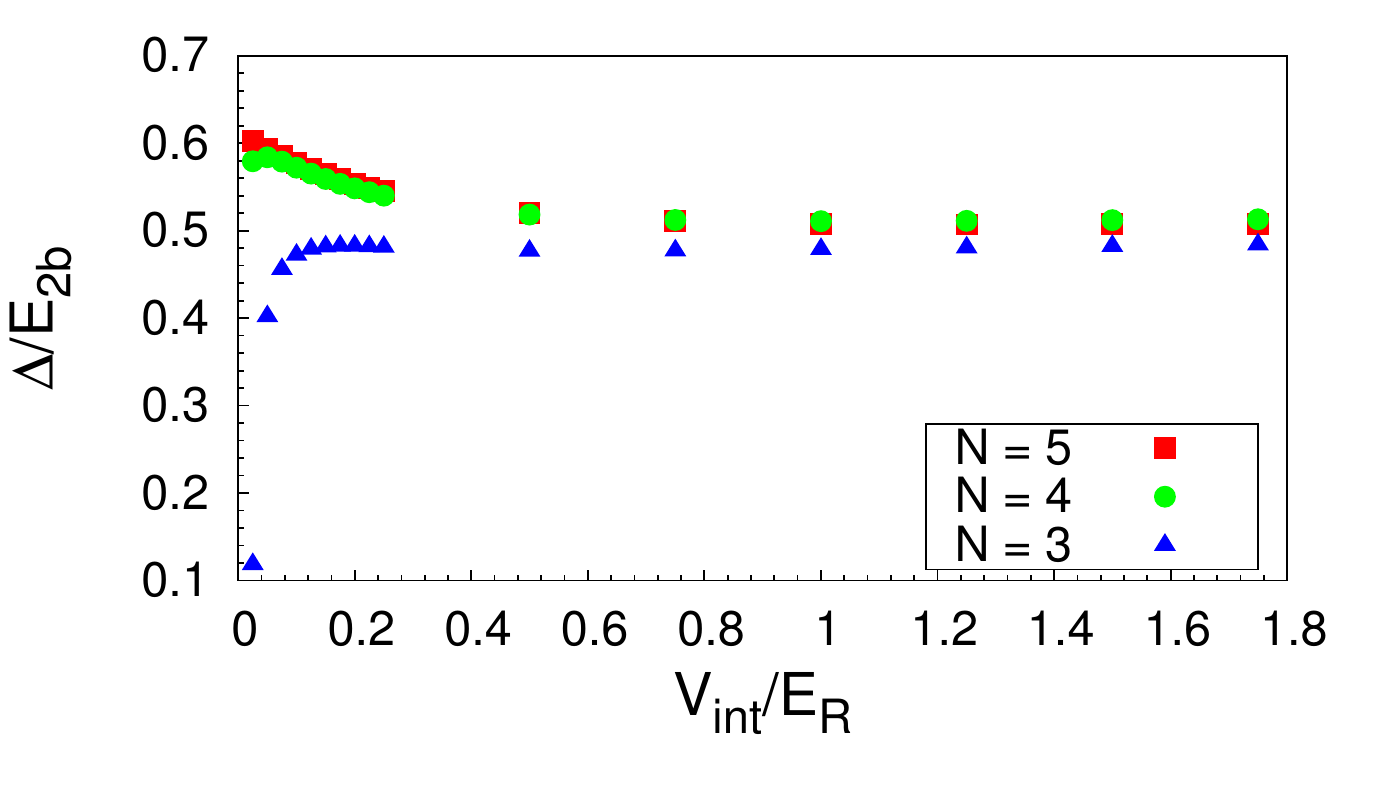}
 \caption{Gap $\Delta$ respective to the two-body interaction energy $E_{2{\rm b}}$ with the two lowest bands taken into account as a function of the interaction strength $V_{\rm int}$ for $N_{\rm b}=8$, $N_x=8$, $N_y=2$ and $V=3E_R$.} 
\label{fig::Laughlin_mixing} 
\end{figure}

In order to investigate the stability of the Laughlin state with respect to the band mixing and the importance of the interaction strength, we implement the projection in the two lowest bands. This model takes into account both the inter-band mixing and the band dispersion, similar to Landau level mixing in the FQH. By varying the interaction strength, one can investigate the role of the mixing. In this model, the two-body interaction energy is not simply given by $V_{\rm int}$, as it was the case for the flat and single band model, since band mixing and band dispersion affect it. In order to make meaningful comparison between these two models, we normalized the energies, for each $V_{\rm int}$ and $N$, with respect to the two-body interaction energy $E_{2{\rm b}}$ of the respective model. The results of this procedure for the Laughlin state with $N_{\rm b}=8$ particles are shown in Fig.~\ref{fig::Laughlin_mixing}. One striking difference with the flat band limit is the absence of a Laughlin state for $N=3$ in the weak interacting regime. This is similar to what was found in the related model of Ref.~\onlinecite{PhysRevLett.110.185301}. This effect is due to the dispersion of the lowest band, which overcomes a very weak interaction, and should also be present for $N=4,5$. However, in these two latter cases the band splitting is much smaller and the data in Fig.~\ref{fig::Laughlin_mixing} does not include the very small interaction strength necessary to see this effect. For the interaction strengths we have studied we find for small interaction that the Laughlin state is realized and we have again $\Delta \sim 0.6 E_{2{\rm b}}$. Moreover, at larger $V_{\rm int}$, the energy gap saturates to a value that seems almost independent of $N$ while it was growing linearly on the flat band model. Remarkably enough, the Laughlin state is always obtained even though the interaction strength is bigger than the band gap. For instance, for $N=5$ the maximum interaction strength we looked at corresponded to $V_{\rm int}/\Delta_{1b} = 1.5$. The fact that the Laughlin state is stable despite a large band mixing is in agreement with previous work on the FQH on a lattice~\cite{Hafezi-PhysRevA.76.023613,Sterdyniak-PhysRevB.86.165314} and for FCI~\cite{sheng-natcommun.2.389,Kourtis-PhysRevLett.112.126806,2014arXiv1407.6985G}.

\subsection{$\nu = 1$: the Moore-Read state}

We now investigate the emergence of the Moore-Read state at $\nu=1$. The MR state is characterized by a three-fold degenerate groundstate on the torus geometry. Strong numerical evidence shows that the MR state can emerge in the FQH for bosons with the hardcore two-body interaction~\cite{PhysRevLett.91.030402,PhysRevLett.87.120405}. It was shown to appear in the Hofstadter model in Ref.~\onlinecite{Sterdyniak-PhysRevB.86.165314} by looking at the entanglement spectrum but the energy gap between the groundstates manifold and the excited states was of the same order than the groundstates splitting. Most of the FCI models do not exhibit signatures for a MR state with a two-body hardcore interaction. However, it can be stabilized using three-body \cite{Bernevig-2012PhysRevB.85.075128,PhysRevB.85.075116} or long-range interactions~\cite{PhysRevB.88.205101}. 

\begin{center} 
\begin{figure}[htb]
\includegraphics[width=0.8\columnwidth]{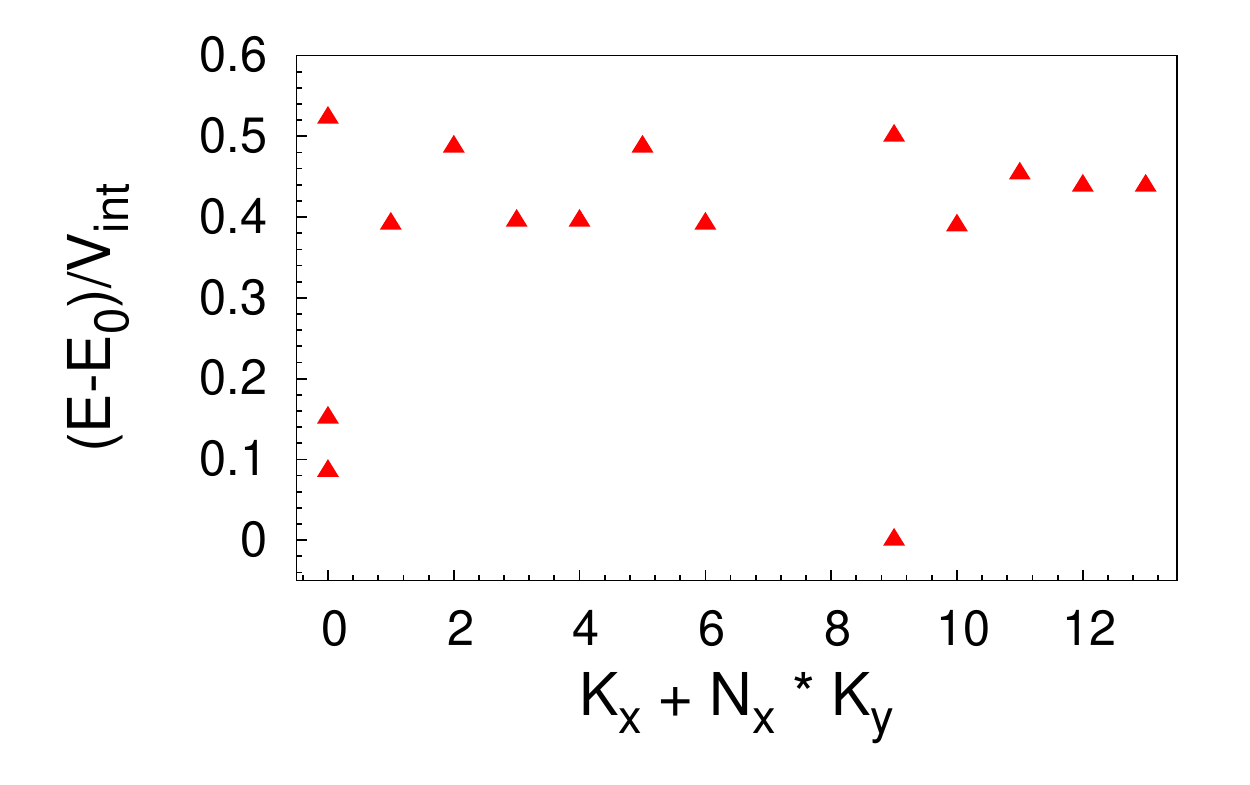}
\caption{Energy spectrum of the two-body interaction at $\nu=1$ for $N=4$, $N_{\rm b}=14$, $N_x=7$, $N_y=2$ and $V=3 E_R$. The energy splitting between the three groundstates is $\delta=0.15 V_{\rm int}$ whereas the energy gap is equal to $\Delta= 0.24 V_{\rm int}$. Energies are shifted by the groundstate energy $E_0$.}
\label{fig::MR_spec} 
\end{figure}
\end{center}

\begin{center} 
\begin{figure}[htb]
\includegraphics[width=0.99\columnwidth]{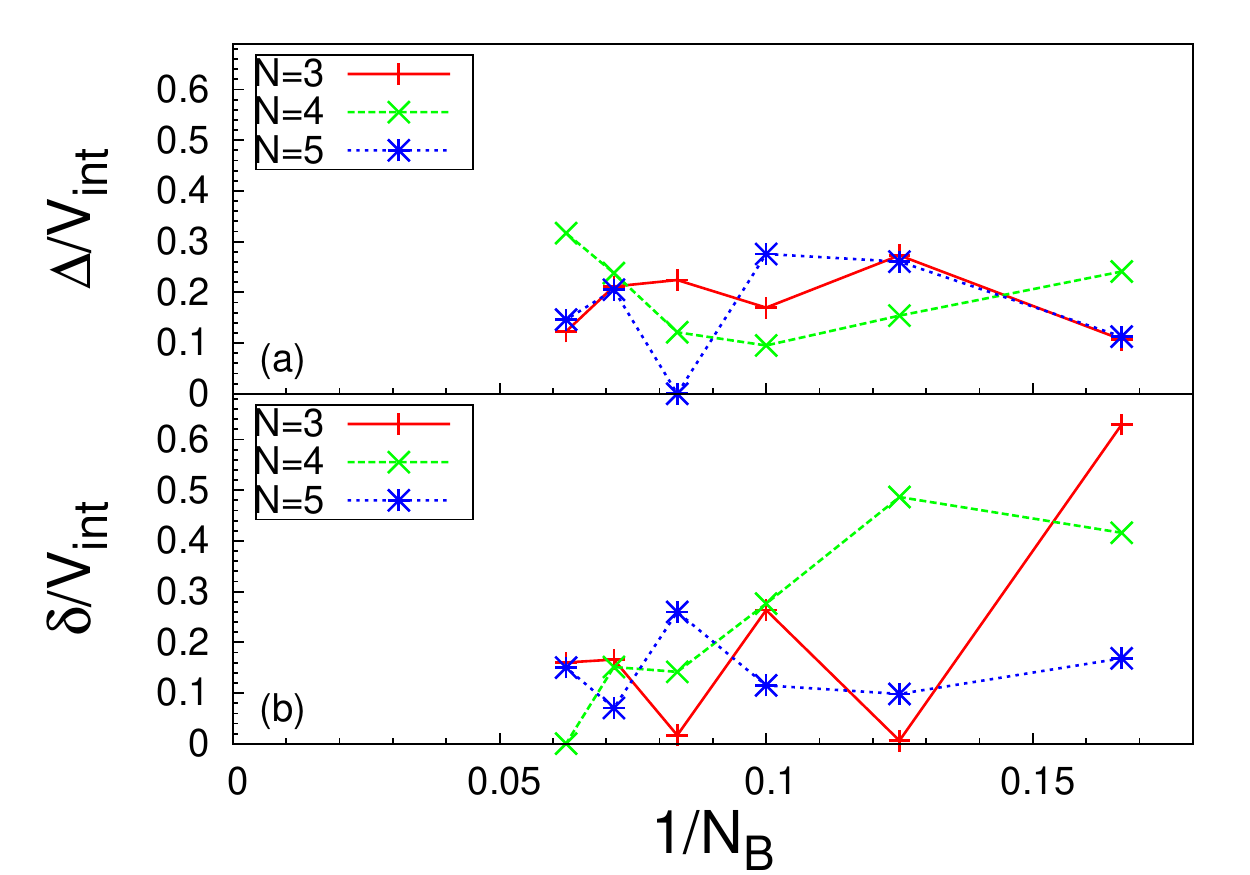}
\caption{Energy gap $\Delta$ (upper panel) and energy splitting $\delta$ (lower panel) as a function of the particle number inverse for different $N$ values with $V=3E_R$.}
\label{fig::MR_gap} 
\end{figure}
\end{center}

Thus it is interesting that numerical evidence for the Moore-Read state was reported in Ref.~\onlinecite{PhysRevLett.110.185301} for weak two-body interactions in an OFL model with $N=3$ which is closely related to the models we investigate here. Our results are qualitatively similar: for most system sizes we find three low energy states in the predicted momentum sectors. However, contrary to the Laughlin case, the energy gap and splitting are of the same order of magnitude, as can be seen in Fig.~\ref{fig::MR_spec} for $N_{\rm b}=14$. The energy gap and splitting are shown in Fig.~\ref{fig::MR_gap}. While the gap and splitting are of the same order of magnitude, the former seems to increase, in average, with the system size while the latter seems to decrease. Note that for the case $N=5$ we find large effects of system geometry: the quasi-degeneracy of groundstates expected for the MR phase is absent for $N_{\rm b}=12$ in a system size $N_x=6$ by $N_y=2$, but present in a system size $N_x=12$ by $N_y=1$, even though the aspect ratio is further from $1$ in the latter case. In general, we find that the MR state is much more sensitive to the aspect ratio than the Laughlin state, which makes any interpolation quite difficult in the system sizes that can be studied numerically. Similar sensitivity of the $\nu=1$ groundstate to boundary conditions ({\it e.g.} the aspect ratio of the torus) appears for the continuum Landau level at small system sizes ($N_b\leq 12$)\cite{PhysRevLett.87.120405} indicative of a competing crystalline phase\cite{Cooper-PhysRevA.75.013627}, but studies on larger systems show a robust Moore-Read state with well-developed gap\cite{cooper2008rapidly}.

\begin{center} 
\begin{figure}[htb]
\includegraphics[width=0.8\columnwidth]{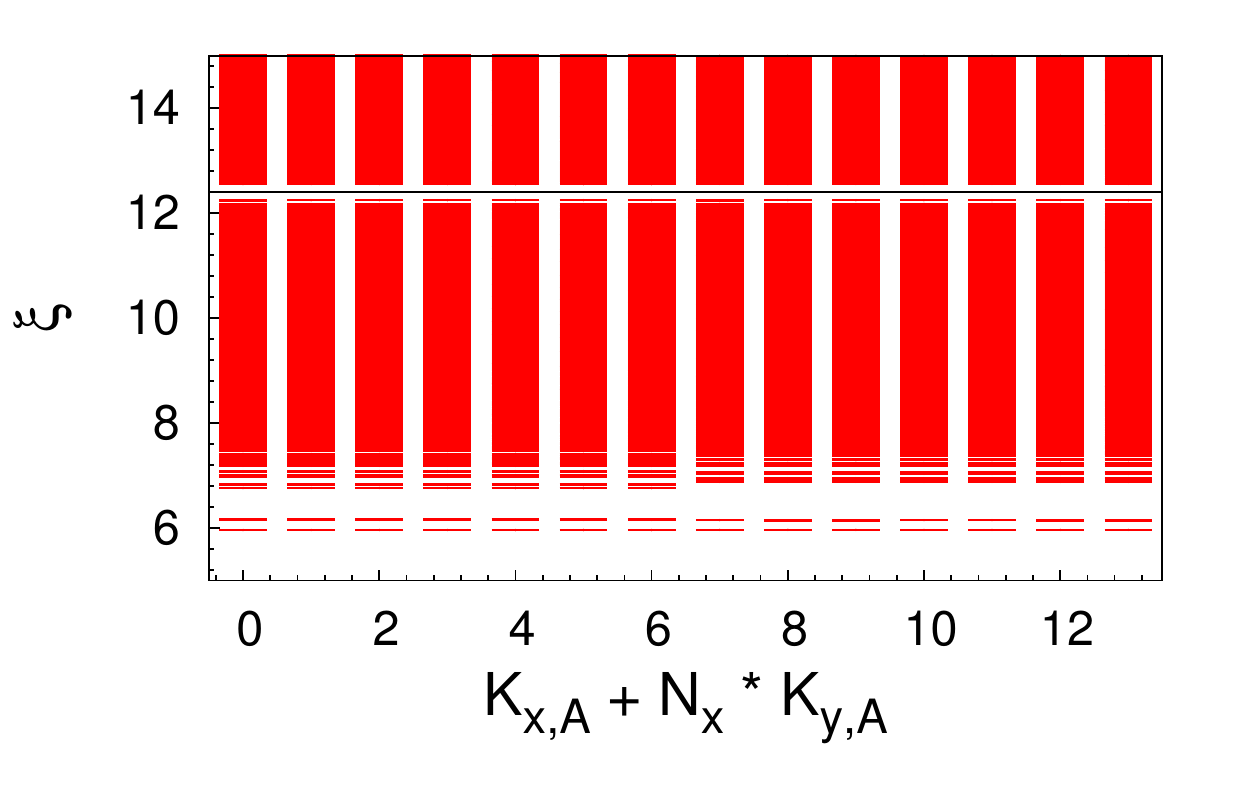}
\caption{Entanglement spectrum of the three lowest energy states for $N=5$, $N_{\rm b}=14$, $N_x=7$, $N_y=2$, $N_A=6$ and $V=3 E_R$. The number of states below the entanglement gap at around $12.5$, is the number of quasiholes states of the MR state.}
\label{fig::MR_PES} 
\end{figure}
\end{center}

\section{Numerical Results for $C>1$}\label{sec:numclt1}

As explained in section \ref{sec:OFL}, OFL models provide a way to tune the Chern number of the lowest band by changing the phase pattern. With $N$ species, $C$ can be set from $0$ to $N-1$. In this section, we investigate the strongly correlated states that can be realized in this model.

\subsection{Two-body spectrum}

\begin{figure}[htb]
\includegraphics[width=0.8\columnwidth]{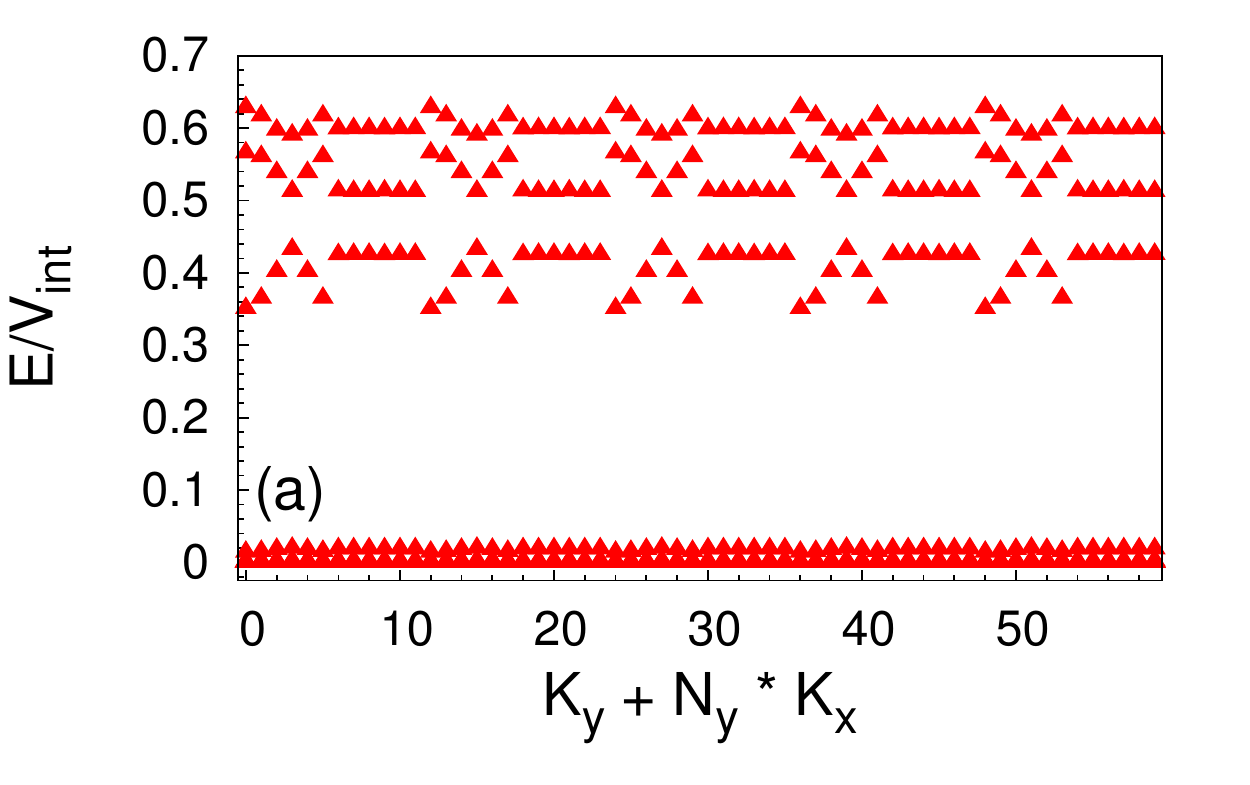}
\includegraphics[width=0.8\columnwidth]{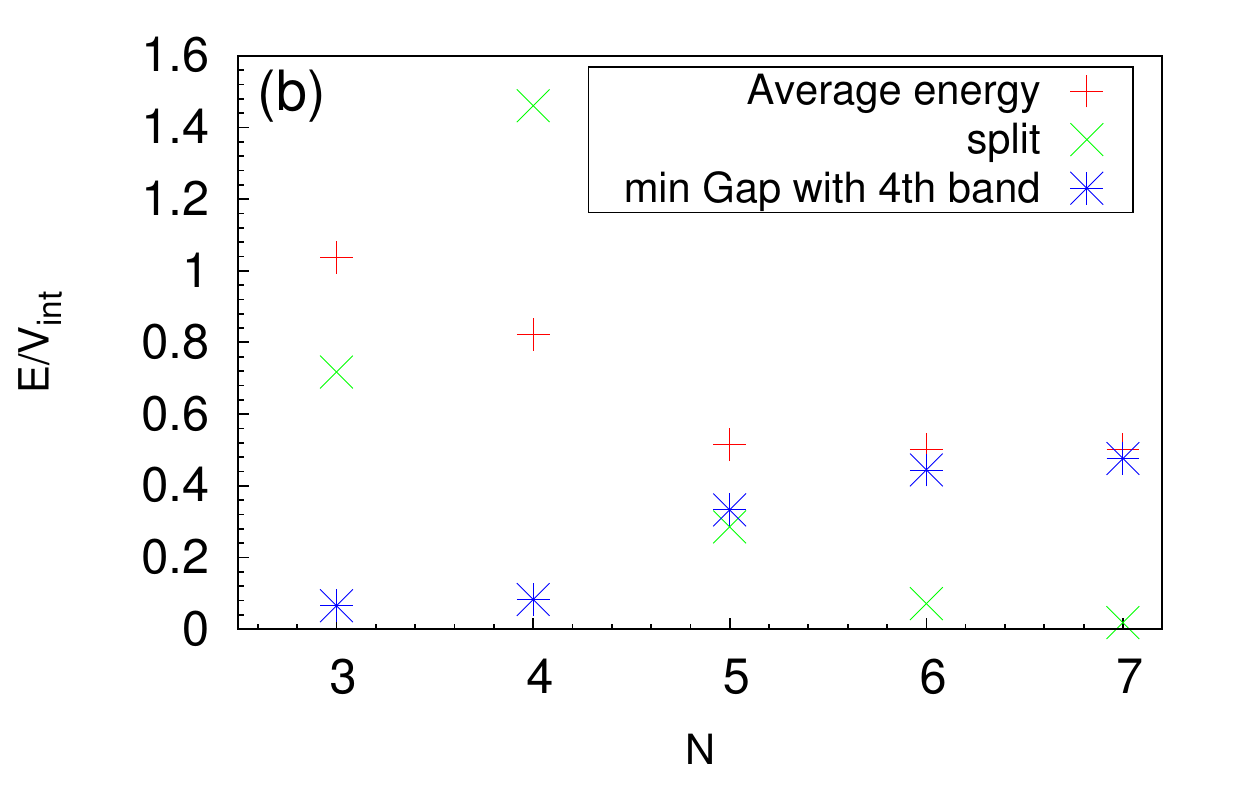}
\caption{\textit{Upper Panel:} Two-body spectrum of the interaction given by Eq.~\ref{eq::int_delta} in the OFL model for $C=2$, $N=5$, $N_x=10$, $N_y=6$ and $V= 3 E_R$. The three branches have average energies around, respectively, $0.4 V_{\rm int}$, $0.5 V_{\rm int}$ and $0.6 V_{\rm int}$ \textit{Lower Panel:} Average energy, energy splitting of the three highest bands and minimum gap to the 4$^\mathrm{th}$ band of the two-body spectra for $C=2$ extrapolated to infinite system.}
\label{fig::higherc_2body}
\end{figure}

As in the $C=1$ case, we start our investigation of the consequence of interaction in these bands by looking at the two-body problem. The two particles spectra for $C=2$ is shown in Fig.~\ref{fig::higherc_2body}. For $C=1$, we obtained (Fig.~\ref{fig::delta_2body}) two branches of non-zero energy states. Those branches where almost dispersionless. Moreover, both the energy splitting between the two branches and the low energy manifold were quickly approaching zero. This allowed us to extract a precise value for the two-body interaction strength in the lowest band and to use that value to normalize the energies. For $C>1$, as can be seen in Fig.~\ref{fig::higherc_2body} in the $C=2$ example, we have $C+1$ branches. We checked that this is also the case when one considers two-component bosons in the lowest Landau level on the torus interacting through spin-independent hardcore interaction. Thus, a natural generalization for $C>1$ of the conjecture of Ref.~\onlinecite{PhysRevLett.111.126802}, would be that ``good'' Chern insulator models for hosting $C>1$ FCI states should have $C+1$ bands separated by a sizeable gap from the other states. Such a rule has consequences on which kind of models should be considered. In the atomic limit, \ie without any tunneling, one can compute the number of non-zero energy two-particle states a given interaction gives rise to. For example, if one considers a model with $N_{\rm s}$ sites per unit cell with on-site interaction, one has only $N_{\rm s}$ two-particle states per unit cell with non-zero energy. Then, obviously, if one needs $C+1$ bands to have a stable FCI states, $N_{\rm s}$ has to be greater than $C+1$ as the projection of the interaction Hamiltonian onto the topological band cannot increase its rank (Since the projection operator is just a matrix multiplication) \cite{2014arXiv1407.0329U}. Such a rule seems to be verified by every model in the literature where this kind of topological state were reported.

In the model we investigate, these bands are much more dispersive than in the $C=1$ case. Indeed, the average energy and the energy splitting between the bands depends more on the number of spin flavor as well as on the system size on both directions. We study in detail the $C=2$ case. In particular, we compute for different systems the average energy and the energy splitting of the three lowest bands, defined as being the difference between the highest and the lowest energy in these bands irrespective of momentum. We extrapolate these values for different $N$ in the limit of infinite system sizes. The results are shown in Fig.~\ref{fig::higherc_2body}b. As expected, the splitting between the band goes to zero as we increase $N$ while the average band energy seems to converge to $ V_{\rm int}$. However, the energy splitting has a non monotonous behavior as its maximum is reached at $N=4$.

\subsection{$\nu = \frac{1}{C+1}$: Halperin-color states}

As it was shown in Ref.~\onlinecite{Wang-2012arXiv1204.1697W} for $C=2$ and in Refs.~\onlinecite{Liu-PhysRevLett.109.186805,PhysRevB.87.205137} for $C>1$, interactions in a fractionally filled band with Chern number $C$ can lead to emergence of Halperin-like bosonic states at filling factor $\nu=\frac{1}{C+1}$. These states are characterized by a $(C+1)$-fold degenerate groundstates. 

\begin{figure}[htb]
\includegraphics[width=0.8\columnwidth]{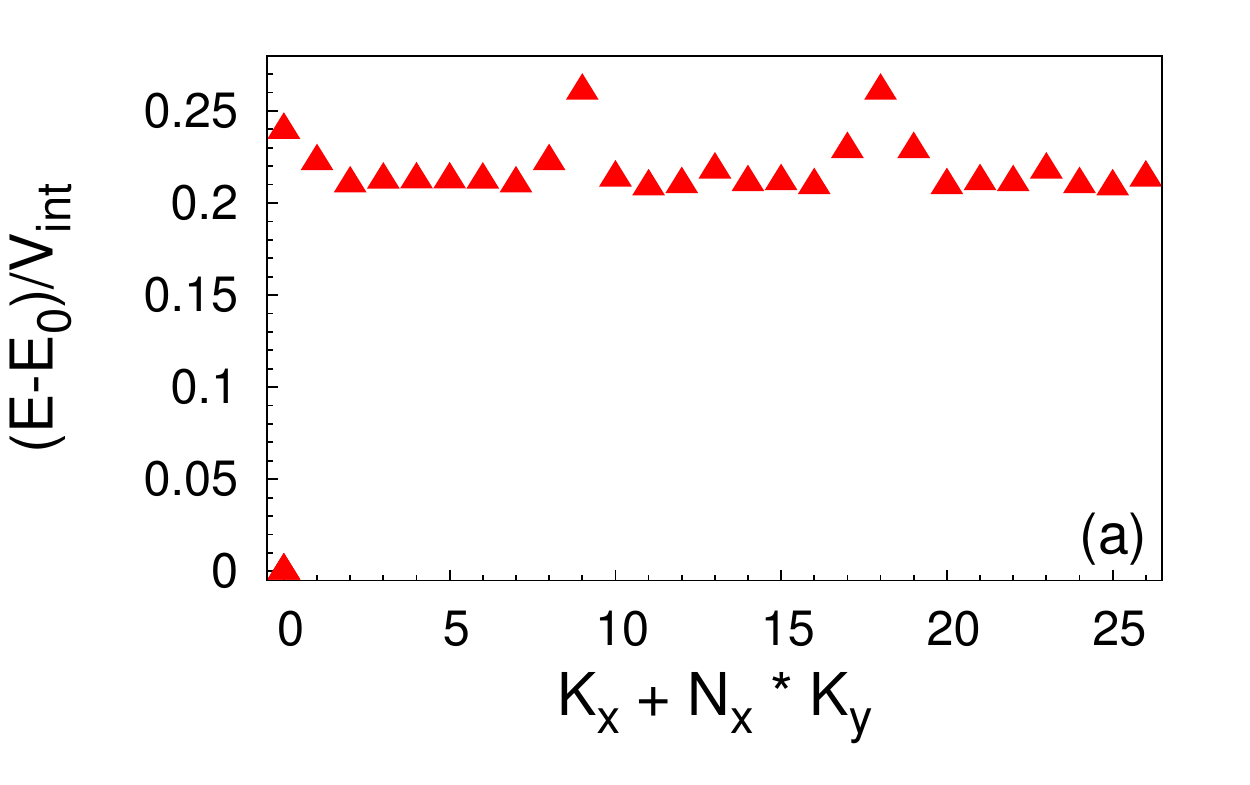}
\includegraphics[width=0.8\columnwidth]{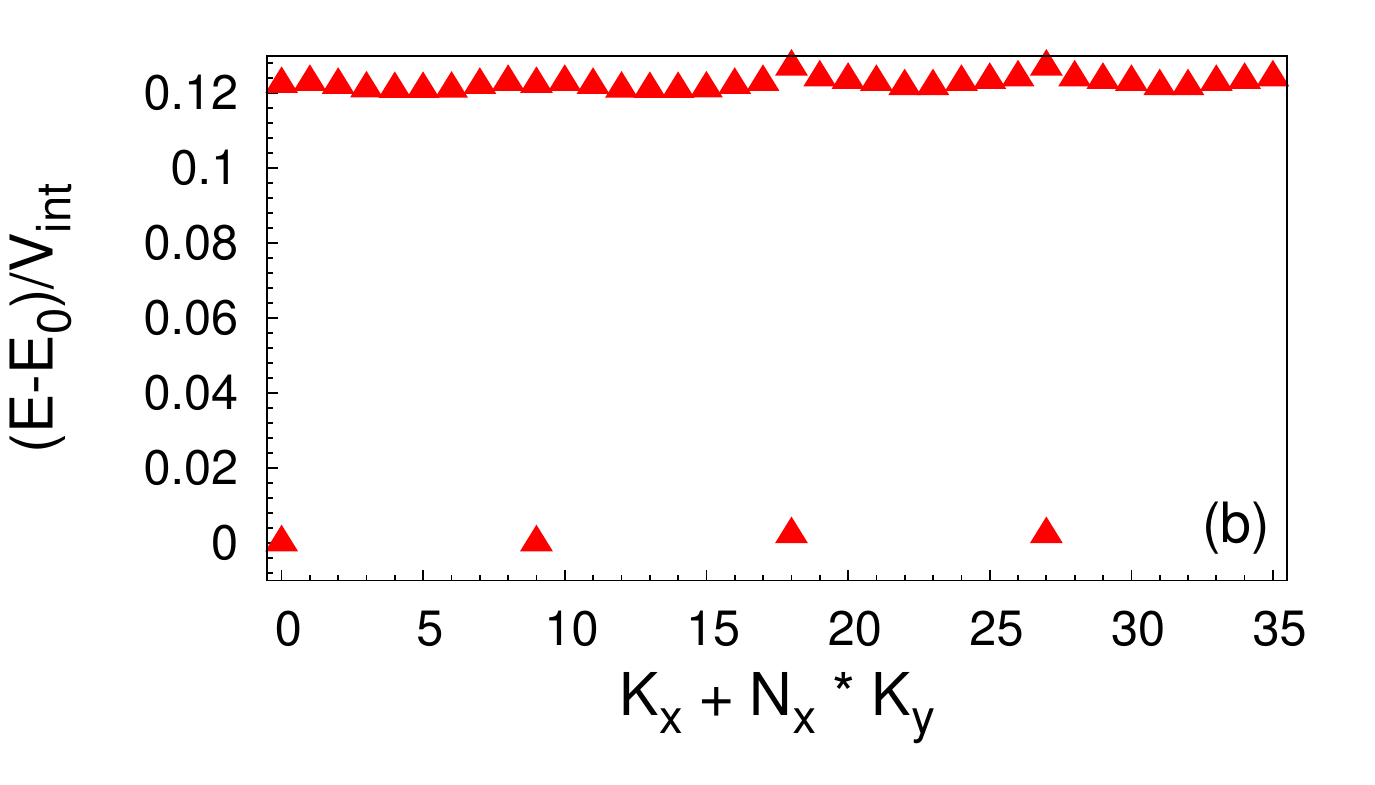}
\caption{\textit{Upper Panel:} Energy spectrum of the two-body interaction for $C=2$ with $N=5$, $N_{\rm b}=9$, $N_x=9$, $N_y=3$ and $V = 3 E_R$. The groundstate in the sector $(0,0)$ is three fold degenerate. The energy splitting between the three groundstates is $\delta=4.6~10^{-5} V_{\rm int}$. The energy gap is equal to $\Delta= 0.21 V_{\rm int}$. \textit{Lower Panel:} Energy spectrum of the two-body interaction for $C=3$ with $N=5$, $N_{\rm b}=9$, $N_x=18$, $N_y=2$ and $V = 3 E_R$. The groundstates are four-fold degenerate and are found in the sectors $(0,0)$, $(0,1)$, $(9,0)$ and $(9,1)$ . The energy splitting between the four groundstates is $\delta=2.2~10^{-3}~V_{\rm int}$. The energy gap is equal to $\Delta= 0.12~V_{\rm int}$. Energies are shifted by the groundstate energy $E_0$.}
\label{fig::Halperin_energyspectra} 
\end{figure}

In the OFL model we studied, we found evidence for the existence of these phases for $C=2$ and $C=3$ with two-body interaction. Examples of the energy spectrum for these two cases are shown in Fig.~\ref{fig::Halperin_energyspectra}. As explained in Sec.~\ref{sec::fci_higherC}, for $C>2$ and $N_A > \lceil\frac{N_{\rm b}}{C}\rceil$, the PES counting of the Halperin-color states are not identical to that of the corresponding $SU(C)$ singlet Halperin states. We found that this is indeed the case for the states we obtained for $C=3$ while, for $C=2$, the PES countings are, as predicted, identical to that of the $(221)$-Halperin state. 

\begin{figure}[htb]
\includegraphics[width=0.99\columnwidth]{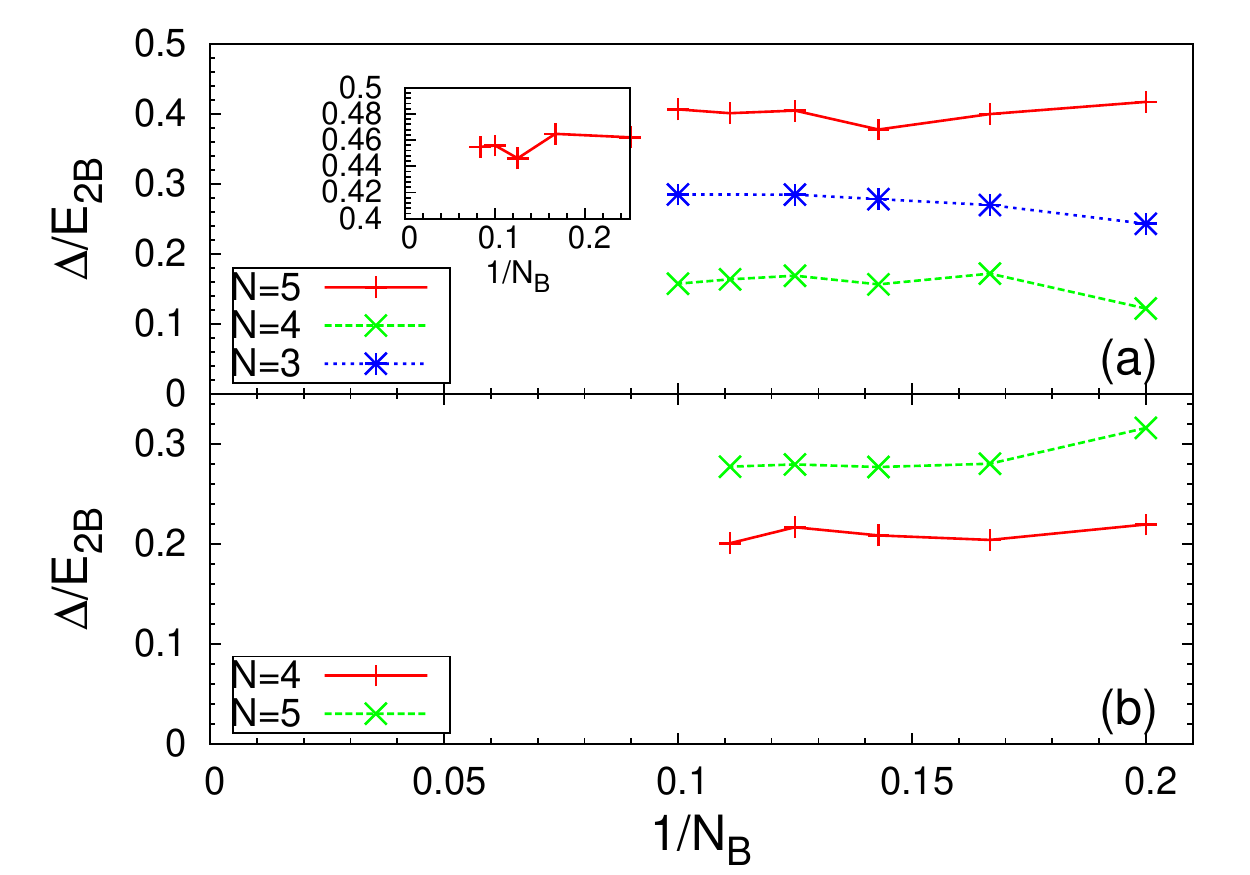}
\caption{\textit{Upper Panel:} Gap as a function of the particle number inverse for different $N$ values with $V=3E_R$ for $C=2$ at $\nu=1/3$. The energy gap are normalized by the two-body interaction energy. The inset shows the gap of the model interaction of the $(221)$-Halperin state on the torus at filling $\nu=\frac{2}{3}$. \textit{Lower Panel:} Gap as a function of the particle number inverse for different $N$ values with $V=3E_R$, $C=3$ at $\nu=1/4$. Since only two system size are accessible for the corresponding Halperin state (namely $N_{\rm b}=6$ and $N_{\rm b}=9$), we cannot provide the gap extrapolation for the FQH on the torus. Still, the gap at $N_{\rm b}=9$ is $\simeq 0.37$ is qualitatively in agreement with what we observe for the OFL model.}
\label{fig::gap_higherC}
\end{figure}

The gap as a function of the inverse particle number, normalized by the two-body interaction energy, is shown in Fig.~\ref{fig::gap_higherC}. Due to the system sizes that can be numerically reached, we will focus on the $C=2$ case. As can be expected from the two-body spectrum, the effect of the number of spin species and of the number of particles on the gap is more important than in the $C=1$ case. It is hence more difficult to give a precise extrapolation in the thermodynamic limit. Notably there is a clear difference between $N=4$ and $N=3,5$ which is reminiscent of the one of the two-body spectrum (Fig.~\ref{fig::higherc_2body}b). We compare these results to the gap we obtained for the FQH problem of spin-$\frac{1}{2}$ bosons on a torus at filling $\nu=2/3$ where $(221)$-Halperin state is realized. The gap convergence is not as good as the one of the model interaction for the $\nu=\frac{1}{2}$ Laughlin state. Nevertheless, the inset of Fig.~\ref{fig::gap_higherC} gives a gap around $\Delta_{(221)} \sim 0.45 V_0$. For $N=3,4$, the gaps that we have obtained in the OFL model are lower than those of the FQH case. For $N=5$, the gap seems to extrapolate to a value closer to $\Delta_{(221)}$. Notice that for the FQH, the gap is only defined when the number of particles is even while the $C=2$ OFL model does not have this restriction. The gap for the $C=2$ OFL model does not exhibit any parity effect. This reinforces the picture that this phase is indeed identical to the one of the $(221)$-Halperin, but with ``twisted'' boundary conditions for $N_x$, $N_y$ odd. 

\begin{figure}[htb]
\includegraphics[width=0.99\columnwidth]{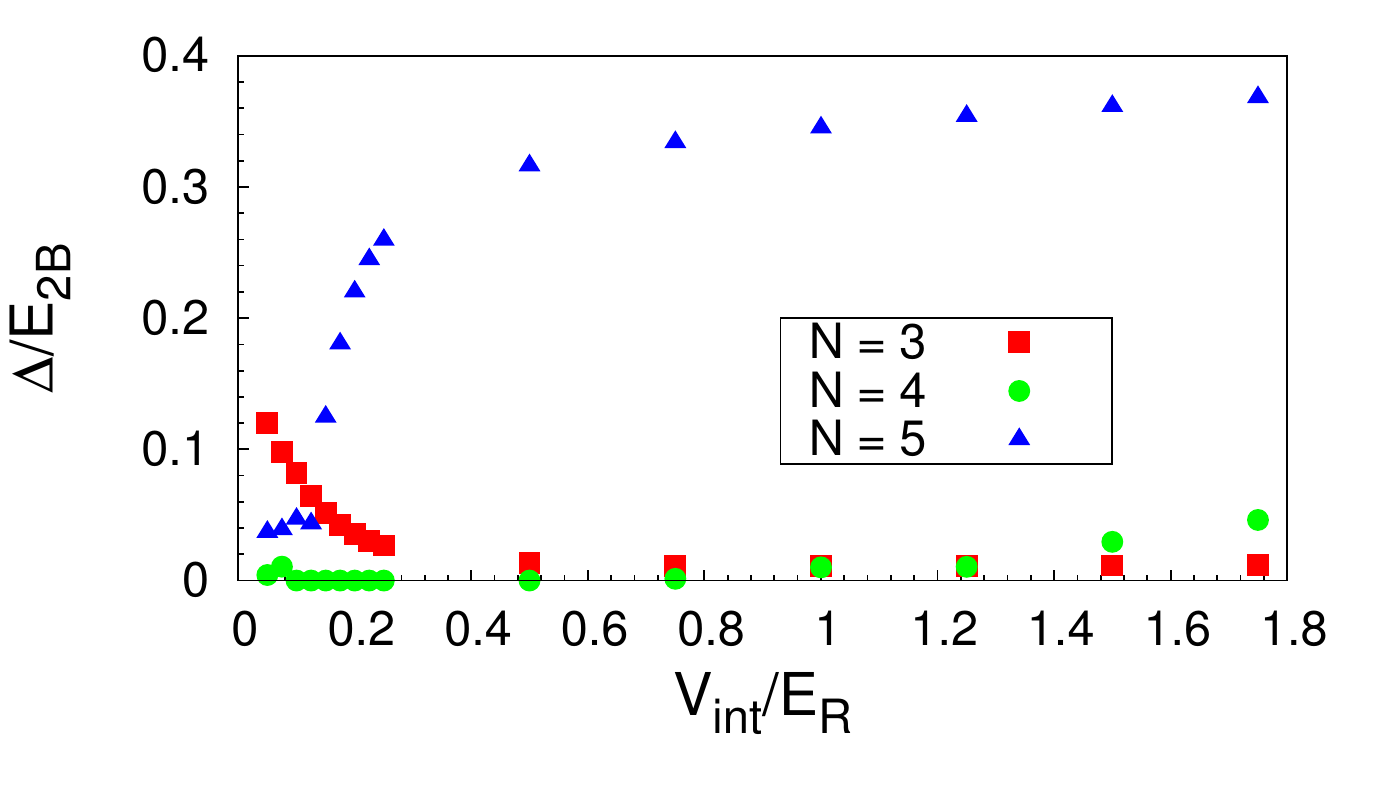}
\caption{Gap as a function of the interaction strength when the two lowest band are taken into account for $N_{\rm b}=7$, $N_x=7$, $N_y=3$ and $V=3E_R$. The energy gap are normalized by the two-body interaction energy computed for each value of $N$ and $V_{\rm int}$.}
\label{fig::gap_higherC_band_mixing}
\end{figure}

We now investigate the effect of the band mixing on the realization of these states. Once again due to technical challenges, we will restrict to the $C=2$ case. The results are shown in Fig.~\ref{fig::gap_higherC_band_mixing}. The behavior changes quite dramatically depending on $N$. For $N=3$, we find that the Halperin states are realized for small interaction but are destroyed by the band mixing. In the $N=4$ case, the system seems unable to support this kind of state. For $N=5$ we find that the state is stable against band mixing but is destroyed by the band dispersion for weak interaction. The value of the interaction strength where the transition $(V_{\rm int} \simeq 0.2 E_R$) occurs is almost an order of magnitude larger than the single particle band width ($\delta=3.3\times 10^{-2} E_R$ as shown in Fig.~\ref{fig:onebody_c_2_c_3}b). For weak interaction, we have indications through the PES that the phase might be a Bose-Einstein condensate. Indeed, we found that the PES has low-lying states whose number is independent of $N_A$~\cite{Sterdyniak-PhysRevB.86.165314}. This is different from band mixing studies for fermionic systems~\cite{2014arXiv1407.6985G} where a metallic phase is observed in this interaction regime. For the situations were FCI states occur, while the gap is smaller than the one extracted in the flat-band approximation, it is still robust even for large interaction and the ground state manifold has the same property as in the flat-band limit. Note that this behavior is insensitive to the particle number parity. The robustness of the Halperin-like phase is similar to that found in section \ref{sec:laughlin} for the Laughlin state in the $C=1$ band. These results suggest that the band mixing would not completely wash out a FQH phase as long as the physical interaction is an implementation of the related model interaction (here the two-body contact interaction). One has to be careful about the conclusion drawn from these calculations, as it is challenging to include more bands in our simulations. Contrary to the tight-binding models of FCIs where the number of bands is finite, OFLs have an infinite number of bands (similar to the situation of Landau levels). In the large interaction limit, a possible scenario is that including more bands will actually decrease the value of the gap, as suggested by our two-band results. We will try to address this open issue in future works.

\begin{figure}[htb]
\includegraphics[width=0.8\columnwidth]{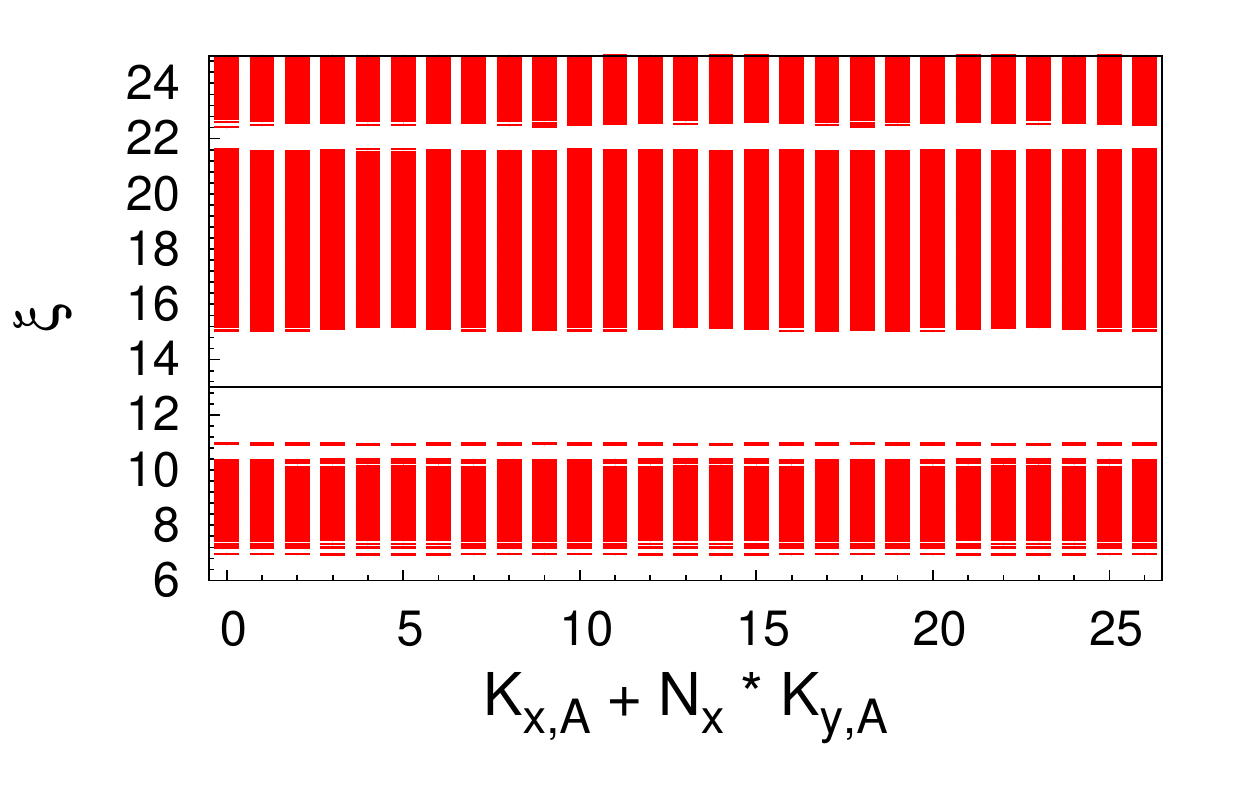}
\caption{Particle Entanglement spectrum of the three groundstates for $N=5$, $N_{\rm b}=9$, $N_x=9$, $N_y=3$, $V=3E_R$ and $N_A=4$. The counting of states below the gap is equal to $(1,2)_2$ spinful counting.}
\label{fig::Halperin} 
\end{figure}

\section{Conclusion}

In this paper, we have investigated the emergence of bosonic fractional quantum Hall states in optical flux lattices. We have quantitatively studied the stability of the Laughlin phase at half filling of the lowest band with Chern number $C=1$ of an OFL. In the flat band approximation, we have numerically shown that the extrapolated neutral gap for this system matches the one of the FQH model interaction for which the Laughlin is the exact densest zero energy state. Such a technique can be applied to any good FCI model to deduce the neutral gap in the thermodynamical limit once the two particle energy scale is known. We have also investigated the effect of band dispersion and band mixing and obtained convincing evidence that the Laughlin phase should stay stable. Beyond the Laughlin state, we have also observed signatures of the Moore-Read state at filling $\nu=1$ and with two-body contact interaction. But the finite size results point toward a much more fragile phase at $\nu=1$ than at $\nu=\frac{1}{2}$.

By changing the Chern number $C$ of the lowest band of the OFL model, we have studied the emergence of Halperin-like phases at filling factor $\nu=\frac{1}{C+1}$ when $C>1$. Similar to the Laughlin case: they show a clear neutral gap whose value can be related to that of the FQH case and they are stable to the band dispersion and band mixing for $N$ large enough and for strong interaction. The robustness to large band mixing of these states whose exact model interaction on the FQH side is just the contact interaction is intriguing. Further studies will try to address this property and especially the effect of the nature and the number of higher energy bands.
 
While flux optical lattices have not yet been realized in a laboratory, unlike tight-binding based Chern insulator model~\cite{2014arXiv1406.7874J} and Harper-Hofstadter Hamiltonian\cite{Aidelsburger-PhysRevLett.111.185301,Miyake-PhysRevLett.111.185302}, we have found that this approach leads to much more stable phases. It therefore represents a promising way to observe FQH states in cold atom systems. Our quantitative studies of the gap, including dispersion and band mixing, provide important information on the temperature scales required for the experimental realization of these phases.

\begin{center}ACKNOWLEDGEMENTS\end{center}
AS thanks Princeton University for generous hosting. AS acknowledges support through the Austrian Science Foundation (FWF) SFB Focus (F40-18). BAB and NR were supported by NSF CAREER DMR-0952428, ONR-N00014-11-1-0635, MURI-130- 6082, Packard Foundation, and Keck grant. NR was supported by ANR-12-BS04-0002-02 and by the Princeton Global Scholarship. NRC was supported by EPSRC Grant EP/J017639/1. This work was supported by the Austrian Ministry of Science BMWF as part of the Konjunkturpaket II of the Focal Point Scientific Computing at the University of Innsbruck. Part of the numerical calculations were performed using the TIGRESS HPC facility at Princeton University.

\bibliography{oflhighc}

\end{document}